\documentstyle[12pt,aaspp4]{article}

\def \etal { et~al.\ }  

\begin{document}

\title{Calibrating the Mixing Length Parameter for a Red Giant Envelope}
\author{S.M. Asida\altaffilmark{1}}
\affil{Steward Observatory, University of Arizona ,Tucson, AZ 85721, USA}
\authoremail{sasida@as.arizona.edu}
\altaffiltext{1}{E-mail: sasida@as.arizona.edu}
\begin{abstract}
Two-dimensional hydrodynamical simulations were made to calibrate the mixing 
length parameter for modeling red giant's convective envelope. As was briefly
reported in \cite{at97}, a comparison of simulations starting with  
models integrated with different values of the mixing length parameter, has
been made. In this paper more results are presented, including tests of the 
spatial resolution and Large Eddy Simulation terms used by the numerical code. 
The consistent value of the mixing length parameter was found to be 1.4, for a
red giant of mass $1.2 M_{\sun}$, core mass of $0.96 M_{\sun}$, luminosity of
$200 L_{\sun}$, and metallicity $Z=0.001$.
\end{abstract}


\section{Introduction}

Convection is a major process taking place in reg giant (RG) envelopes. In a
previous letter (\cite{at97}, hereafter AT97), results of a study of this 
subject were reported briefly. In the present paper, more results of this study
 are presented. 

Throughout the paper, the reference one-dimensional (1D) model of convection is 
the mixing length theory (MLT, \cite{bv58}), according to the prescription of 
\cite{cg68}. Laboratory experiments, observations of the outer layers of
the solar convective zone, and theoretical understanding have demonstrated 
that convection is much more complicated than MLT assumes. 

Analysis of 
experiments of convection in a box, in which a temperature gradient was 
established by using Helium at about 5K and controlling the  boundary 
conditions, have shown that convective flow is non-local with defined narrow 
streams going from the top down (and vice versa), i.e. the existence of a 
downstream at a given depth at a given time depends upon the existence of a 
downstream in a higher point at previous time and not upon local parameters only. 
This phenomenon is known as hard turbulence (see \cite{hel87,cas89,wl92,tli93}).

Observation of the outer layers of the sun have shown the existence of two 
dimensional cells of upstreams , in a variety of sizes and shapes, with the 
boundaries of these cells being the downstreams, i.e. the flow is not symmetric
(see \cite{ns95}).

During the years, many studies were made in order to overcome the 
simplifications of MLT. Several non-local and non-instantaneous theories were 
proposed (\cite{spg63,ulr70,ulr76,stl82,gw92,gn93,grs93,grs96,xng97,cnt97,cd98}). 
A common feature in these theories is using turbulence theory
techniques in which the hydrodynamical quantities at each point are constructed
of two components: a constant value and a varying value. Equations are
formulated for each of the varying quantities which are then used to determine
the convective velocity and flux. These equations include creation and 
diffusion terms and thus constitute a set of coupled partial difference 
equations. Maybe because of this complexity, there is no wide usage of these 
theories. Xiong \etal (1998a,b) have used one of these theories to compute 
linear pulsation times of RR Lyrae and Long Period Variables. \cite{bch99}
are using a simplified 1D model of turbulent energy diffusion and
convection with eight free parameters to describe Cepheid and RR Lyrae variables.

\cite{cm91} used turbulence models to construct an alternative 
model for MLT which includes a full spectrum of eddy sizes (see also 
\cite{cgm96}). This model is still local and instantaneous, but is easy to 
include in astrophysical codes. It was found that this model yields higher 
values of convective flux than MLT at efficient convection regimes, and lower
values of convective flux at inefficient convection regimes. This model was 
found to be preferable to MLT in many studies (see \cite{cnt96,cc98}). 
These studies included red giant evolution (\cite{sc95,sc97}), 
Heliosesmology (\cite{bsu94}) and calculations of uvby colors of A F and G
 stars (\cite{sk97}). The superiority of this model, for example in 
modeling red giant evolution, was the usage of a constant parameter for 
various masses, while when using MLT there was a need to change the mixing length 
parameter with mass. However, \cite{ldw99} have shown that a varying calibration 
is needed for this model too.

Alternatives to the usual mixing length used in MLT (which is a free parameter
times the pressure scale height) were suggested. These included using the 
density scale height (\cite{sw61}) and the distance from the boundary of 
the convection zone as in \cite{cm91}.

For a correct modeling of red giant variables, the lack of time dependence 
and the locality of MLT 
are major disadvantages. The convection velocities, as predicted by MLT, and 
the radial velocities associated with pulsation are of the same order of 
magnitude (of $10 \case{km}{sec}$). Thus, interaction of the two phenomena is 
expected.

Ad hoc prescriptions were made in order to take into account time dependence 
in dynamical calculations with MLT (for example, pulsation calculation of 
Miras, see \cite{tt80}). In these calculations, the convective flux at a given
 time is determined by using MLT and the flux at a previous time assuming a 
typical convective time scale estimated by convection velocity. In other 
calculations, like evolution of massive stars, a non locality is presented to 
MLT in particular when treating penetration (or overshoot) of convection to 
stable neighboring zones. A common assumption is that this overshoot region is 
proportional to the mixing length at the edge of the convection region. These 
methods are obviously not rigorous and not intended to describe these 
processes precisely (for example see \cite{mc94}.

Being mainly a hydrodynamical phenomenon, an appealing method for modeling 
convection is by using numerical multidimensional simulations. In this 
approach, no explicit model of convection is needed, and thus no mixing length
parameter is assumed. Time dependence and non locality are naturally present.
However, convection is multidimensional and the flow has many length scales 
and time scales, that might be smaller by far than other typical scales of the
problem. This is why such calculations are not suitable for all problems with
convection.

Numerical simulations of the hard turbulence experiments mentioned above 
where made (Werne 1993,1994, \cite{ker96}). These calculations reproduced the 
typical narrow streams of the experiments, and revealed that the origin of the
flow was the thin boundary layers at the bottom and at the top of the 
experimental box.

Many studies were made in order to understand general features of convective 
flow (\cite{hrl84,cs86,mlg90,ctn90,hm91,ctn91,cg92,sc93,rt93,xt93,pw94}). 
Typically, these studies assumed 
simplified physical assumptions, such as ideal gas equation of state, constant 
opacity and constant gravitation. The systems were confined in a box with an 
initial unstable polytropic profile. Boundary conditions were periodic on the 
sides of the box, no flow was assumed in and out of the bottom and top of the 
box, a constant temperature gradient at the bottom and a constant temperature 
at the top. Some of the simulations assumed planar symmetry using 2D codes, 
while other used three-dimensional (3D) codes.

Summary of some the results of these studies is: there is an asymmetry in the 
flow with narrow rapid downstreams and wider upstreams and the 3D 
picture resembles the observations of the sun with cells of upstreams; the 
flow in the wider upstreams is somewhat more turbulent; the size of the cells 
increases with depth so downstreams merge. This picture is revealed when using
high resolution, while low resolution simulations show laminar flow with big 
convection cells. The internal energy flux in the downstreams is relatively 
large, but a large kinetic energy flux with opposite sign almost cancels the 
net energy flux in the downstreams. Near the outer boundary, convection might 
be supersonic.

Other "simulation in a box" studies were dealing with invoking of p-modes in 
the convection zone (\cite{bgd93}), producing g-modes below (and above) 
the convection region (\cite{hrl86}) and penetration of convection to 
neighboring layers (\cite{hrl94} Singh \etal 1994,1995).

Attempts have been made to simulate the outer layers of the solar convection 
region (Stein \& Nordlund 1989,1998, \cite{kim95}). Since the computation box in 
these simulations included only a fraction of the convection region, 
penetrative boundary conditions were applied at the bottom of the computation 
box. The outer layers which were included in the simulations and in which a 
superadiabatic gradient exist, are the most sensitive to the convection model. 
These simulations reproduced the granulation pattern of convective streams of 
solar convection, as well as the heat flux and spectrum 
(\cite{spr90,ns95,nrd96,kim96,abt97,dmr97}). 
Comparison of various measurements form the 
simulations (such as maximal temperature gradient near the outer boundary) 
enables calibration of the MLT parameter. \cite{ldw99} and \cite{fls99} have 
published 
such calibration for various models of solar-type stars. They found that MLT 
parameter is  1.2 to 1.8 for starts with effective temperatures of 7100K to 
4300K (with dependency upon g - gravity and metallicity as well). They also 
investigated the \cite{cm91} model and found it needs varying calibration as well.

Simulations of convection in ZZ-Ceti type white dwarfs were done by 
\cite{ldw94}. They found that different calibration models yield different MLT 
parameter values: 1.5 according to one measurement, and 4 according to another.

Two-dimensional (2D) calculations of RR-Lyra were performed by Deupree (1975,1977
,1985); the (very) low resolution simulations of the narrow convection region in 
these variables stars have shown that convection has a damping effect upon 
pulsation, and vice versa. In these stars, there is little difference between
a radiative model and a convective model, due to the narrow convection region. 
Thus, the relaxation time in the 2D simulations is relatively 
small, and might be calculated. This is not true for RG in which the 
convection region is wide, and that is why simulating convection in RG's 
envelopes is much more difficult.

The above discussion, as well as the topic of this paper, concerns convective 
envelopes, in which the convection is not efficient enough to enforce an almost 
adiabatic profile throughout the convective region. Multidimensional studies of 
other astrophysical cases have been published. This includes core convection 
(\cite{dpr98}), shell burning (\cite{arn94}, Bazan \& Arnett 1994,1998), 
Nova bursts 
(\cite{gl95,gls97,krc98}), core collapse (see \cite{mzc98} and ref. therein), 
and accretion disks (for example \cite{sb96}). In these studies, some of 
the interesting questions are: the amount of 
mixing and penetration due to convective streams, nuclear energy sources and 
sinks in the convection region, the coupling of convection and rotation.

An interesting question in the context of multidimensional simulations of 
convection, is the validity of 2D calculations. While it is obvious that 3D 
simulations are preferable, it is not clear what are the quantitative differences 
between 2D and 3D results. In many 2D studies of convection, there are large
convective cells (eddies) that occupy the whole convective region (see for 
example \cite{hrl84}), this phenomena is not seen in 3D simulations. However, it 
does not mean that there are no small size eddies in 2D simulations, when increasing
spatial resolution large eddies tend to brake up to small eddies (see \cite{pw94}),
and even low resolution 2D simulations have small eddies (see \cite{chn82} and
discussion in \cite{cs86}). The qualitative features of convective flow, revealed by
3D simulations, are present in 2D simulations as well (see \cite{sn89} and ref. 
therein). There are few quantitative measurements of the differences between 2D and
3D simulations: in a study of convection in proto-neutron star, \cite{mj97} have 
shown that the importance of small scale features, and the kinetic energies are 
different when comparing 3D to 2D simulations; \cite{krc99} have simulated novae
in 3D and compared the results to their 2D simulations (\cite{krc98}), they found
the difference in the spectrum of eddy sizes to have an influence on the amount 
of overshoot and mixing, thus yielding a different answer to the question of the 
existence of thermonuclear runaway; in a study of convection in solar type 
stars, which is more comparable to this paper, \cite{ldw99} have results of first 
3D calculations, these results indicate little difference between the calibrated 
mixing length parameter from 3D simulations, relative to the calibrated mixing 
length parameter from 2D simulations.

Modeling the convection region of RG envelopes is necessary when studying 
phenomena related to these stars. For example, consider the observational 
relation of the pulsation time to the luminosities of Mirae: there is ambiguity 
concerning the pulsation mode (is it the fundamental mode or the first mode?).
The period in the first mode is about half of the period in the
 fundamental mode, however changing the mixing length parameter from 1 to 1.5 
is enough to change the calculated period by a factor of 2 
(\cite{yt98}). The value of the mixing length parameter affects the 
calculated age of globular clusters (see \cite{fs99} and ref. therein).

Due to the importance of modeling convection in studying RG's, and the lack of 
previous hydrodynamical calculations of this phenomena, I focused on 2D
simulation of RG's envelope convection. In \S 2 I describe the
code used for the simulations. Summary of the results published in AT97 is
presented in \S 3 and more results of the AT97 simulations are summarized in
\S 4. A survey of the importance of the various parameters of the simulation
techniques is presented in \S 5 and \S 6. A short summary of this research is 
given in \S 7.

\section{The numerical scheme}

As was described in AT97, the hydrodynamical code VULCAN (\cite{lvn93}) was
used for this study, after utilizing several modifications to the code. This 
version of the code solves the hydrodynamic equations 
and radiative diffusion equations 
by splitting each time step into hydrodynamic time step followed by a radiative 
diffusion time step. Both steps are solved implicitly to have an accurate and  
stable solution. The code uses an Arbitrary Lagrangian Eulerian (ALE) scheme in 
which each Lagrangian step is followed by a relaxation of the numerical grid. In 
this relaxation stage, the grid is modified to eliminate distortion and to allow
calculation of flow with vorticity. A quasi 1D Lagrangian relaxation was used, 
in which each radial row of cells kept its mass.
This  was done for each row of cells by calculating an average outer 
radius, so that the sum of the mass fluxes 
through it would be zero. Using this method, one was able to follow 
the average radial expansion or shrinking of the envelope. The mass, momentum and 
energy fluxes in the relaxation
 stage were calculated by a second order donor scheme, in which the gradiants of 
these variables were used to estimate the fluxes at the boundaries of the cells 
(according to \cite{vlr79}). . 

Like other numerical studies of convection (\cite{cs86,cg92,ldw94,kim95}), 
I implemented subscale viscosity and 
diffusion terms using the \cite{smg63} model. The inclusion of these terms in 
the equations solved by the code should compensate for the use of a finite 
number of 
cells which is much less than needed when simulating turbulent flow if one 
wishes to simulate the effects of all scales up to the scale of physical 
dissipation. In the simple Smagorinsky model these terms are the usual 
viscosity terms with a coefficient which depends upon derivatives of the 
velocities and a typical length scale of the numerical grid. When there is a 
scalar field that is characteristic of the fluid, one should add a diffusion 
term of this scalar field with a coefficient which is proportional to the 
viscosity coefficient. In convective flow, this scalar is the entropy 
(\cite{cs86}, Ludwig \etal 1994, \cite{kim95}) and thus entropy diffusion was 
added to the code. The two parameters in these Large Eddies Simulation (LES) 
terms were the Smagorinsky 
coefficient (which multiplies the viscosity coefficient) and the Prandtl number 
(which is the ratio of the viscosity coefficient to the diffusion coefficient). 
Both the viscosity and the entropy diffusion were added as explicit terms since 
they were relatively small.

The inner boundary in the simulations was located deeper than the convection 
zone to minimize the effects of not including the entire envelope. However, in 
preliminary simulations, hydrodynamical waves were invoked below the convection 
zone, and there was a convective flux near the inner boundary. By comparing 
simulations in which the inner boundary was placed at different depths, it was 
found that this convective flux was not physical because it was always located 
near the inner boundary. Realizing this was due to an improper boundary condition 
(in which only the radial velocities were set to zero for points on the inner 
boundary), I changed the condition and set both components of the velocities at 
the inner boundary to zero and allowed  only radial velocities at the lowest 
three rows of points. With this new boundary condition, the flux disappeared. 
The inner boundary condition for the radiative diffusion was of a constant 
entering flux equal to the luminosity of the static model (i.e. equivalent to 
$200 L_{\sun}$).

For the hydrodynamic step a constant (usually zero) outer pressure was used,
without limiting the direction of the velocities at the outer boundary. However,
 at the mesh relaxation stage, an average radius was calculated for the outer 
boundary so that the mass flux through the boundary was zero. The remapping fluxes
(i.e. the fluxes at the mesh relaxation stage), through this boundary were 
calculated
using the values of the bundary cells (i.e., using an acceptor scheme for the 
incoming flow). For the radiative diffusion, an outgoing 
flux was calculated using $ \sigma T^4 $ where T is the temperature of the 
boundary cell and $\sigma$ is the Stefan constant (i.e., assuming it to be at
optical depth of about two thirds).

For the sides of the computational sector either reflective boundary conditions, 
or periodic boundary conditions were used. When using the reflective boundary
 conditions, all fluxes through these boundaries were set to be zero, and thus 
only downstreams or upstreams existed on the boundaries. This was different when 
using periodic boundary condition, for which each boundary cell has an 
artificial neighbor at the other side of the sector (though, since I assumed 
cylindrical symmetry, there is no direct physical meaning to periodic boundary 
conditions).

In order to analyze the results of the simulations I used printouts of time 
averaged energy fluxes for each radial layer. The fluxes included radiative 
flux as was calculated by the radiative diffusion step, remapping fluxes of 
internal energy and kinetic energy calculated by the grid relaxation stage, and 
hydrodynamical fluxes that can be calculated using 
   $ \case{d \epsilon}{dt} = -\case{\nabla \cdot  (p\vec{u})}{\rho} $
where $\epsilon$ is the total energy density, $p$ is the pressure, $\vec{u}$ is
 the velocity vector and $\rho$ is the mass density. The convective flux is the
 sum of the remapping and the hydrodynamical fluxes.

Both the radiative and the remapping fluxes were calculated explicitly in the 
simulations and had only to be time averaged. However, the hydrodynamical 
fluxes were intrinsic and thus had to be estimated. I used two methods: 
a. explicitly using the above formula, and b. by energy balance of each radial 
layer in the hydrodynamical step (i.e. the flux from one layer to the next in 
each time step is equal to the change in total energy of the layer during the 
time step minus the flux from the previous layer). Both methods had 
inconsistencies: the first uses a formula which is not one of the formulae that 
control the simulation and the second is correct only if full conservation of 
energy exists in the simulation. Additional discussion of this issue appears in 
\S 4, where a flux plot is presented.

As a test case for the simulation techniques I reproduced the results of a
``convection in a box'' study
(published by \cite{hrl86}). The configuration for these simulations was 
of two stable layers with an unstable layer in between. The velocity field 
and the averaged fluxes in the simulations were similar to the published 
existing results.

In addition to the 2D hydrodynamic code, two 1D codes were used: a static code
(see \cite{tt78} for description), and a hydrodynamic code (\cite{tt79}). The static 
code was used to integrate initial models of the envelope, and the dynamic code 
was used to follow the hydrodynamic evolution of the models for comparison with 
the 2D results (as explained in the following sections). Both codes use 
\cite{cg68} prescription of MLT to calculate convective flux. The equation of 
state and opacities used by these 1D codes are identical to those used by the 2D
code. The equation of state is based on tabular form and includes sum of 
components for electrons, ions, molecules of hydrogen and helium and radiation (see
\cite{tt78,tt80} for details). Opacities are calculated according to \cite{cs70}
with water molecules contribution according to \cite{pa69} (taken from \cite{tt80}).

\section{Summary of previous published results}

The study was focused on models of an envelope of a red giant with total mass of 
$1.2 M_{\sun}$ core mass of $0.96 M_{\sun}$, luminosity of $200 L_{\sun}$ and 
metallicity of $Z=0.001$. The models were 
integrated using the Mixing Length Theory with the mixing length proportional to 
the pressure scale height. Five models were integrated using different values of 
the proportion coefficient (-mixing length parameter) $\lambda$: $0$ (i.e. no 
convection), $\onehalf$, $1$, $1\onehalf$, and $2$. These five models had 
different total energy as well as different energetic structure as an outcome 
of different energy transport efficiency, as was dictated by the value of the 
mixing length parameter. The aim of the study was to find  which of these five 
models was more consistent with the results of  2D dynamical 
simulations.

The final configuration reached by the end of 2D dynamical 
simulations of each of these five models should be the same, since they all 
correspond to the same physical envelope (the same mass, luminosity and 
structure below the convection zone). The difference in the simulations is in 
the relaxation time needed for the different models to adjust their total energy
 and structure and to reach thermal equilibrium. Due to the long thermal time 
scale of these envelopes,and to computational limitations, this relaxation phase
 could not be fully simulated, however, as was demonstrated in figure 1, there 
was enough data in the beginning of the simulations to reach conclusions about 
the preferred model.

\begin{figure}
\plotone{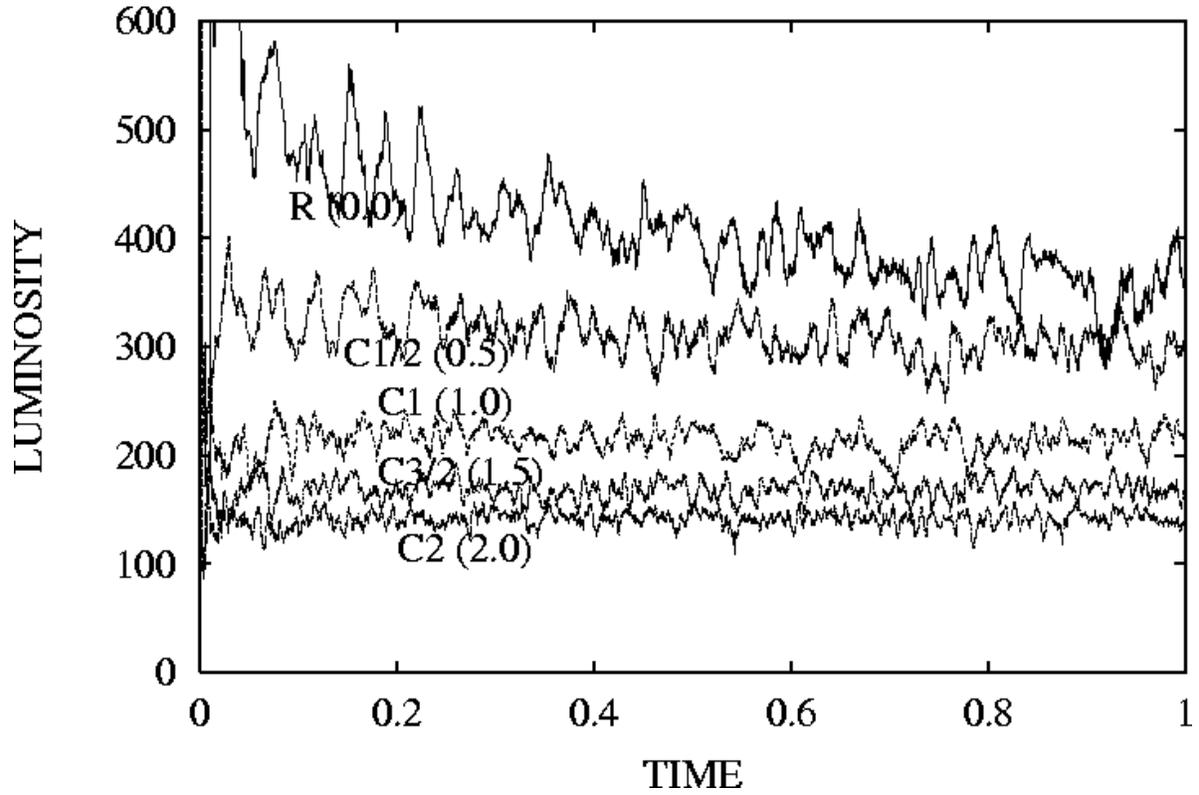}
\caption{Luminosities as a function of time in the 2D simulations 
of different initial models. The $\lambda$ used to integrate each initial model is 
indicated. Luminosities are in $L_{\sun}$ and the time is in years.\label{fig1}}
\end{figure}

\placefigure{fig1}

In this figure the luminosity as a function of time for each of the 
five simulations was presented. The most obvious result in this figure is that 
in the simulations that started from models with $\lambda < 1$ the envelope 
lost energy while for simulations started from models with $\lambda > 1$ 
the envelope gained energy. That was the main reason for the conclusion 
that the 1D model integrated with $\lambda = 1$ was the most
 consistent with the 2D simulations.

A consistency check of this result was done by using 1D dynamic 
simulation with $\lambda = 1$ MLT starting from the above different initial
 models. It was found out that the 1D simulations mimic the 2D
simulations, i.e. the luminosity (and radius) as a 
function of time for the different initial models looked similar. A more 
precise analysis of the results revealed that the final configuration of the 
{\it dynamical} simulations (both one and 2D) had slightly less total 
energy than the {\it static} 1D model of $\lambda = 1$.

\section{More results}

A typical velocity field in the simulations is presented in figure 2. One can 
identify narrow downstreams and wider upstreams in the flow. The length scale 
of the streams increased with depth (as well as the pressure scale height). The
 existence of hydrodynamical waves below the convection zone is apparent.

\begin{figure}
\plotone{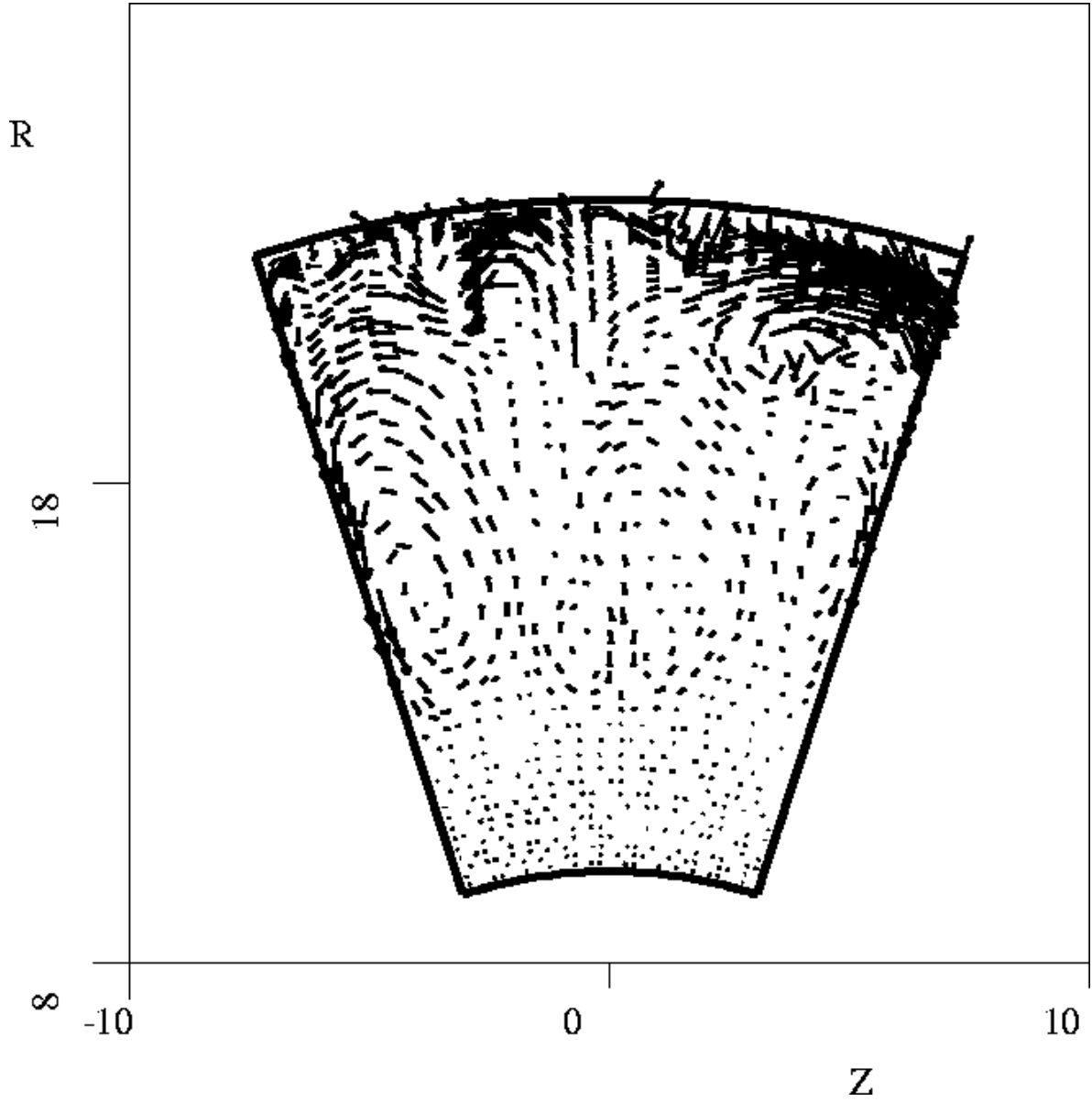}
\caption{The velocity field in the $\lambda = 1$ standard simulation. Z is the 
rotation axis and R is the distance from it in $R_{\sun}$,
 the largest arrows correspond to velocity of $10^{6} \case{cm}{s}$ \label{fig2}}
\end{figure}

\placefigure{fig2}

As mentioned before, by comparison of the five simulations a conclusion could be
made that the $\lambda = 1$ model was the most consistent with the final 2D
configuration, however from figure 3 it is clear that this model 
was not in a full thermal equilibrium. In this figure,the (time) averaged fluxes 
in the 2D simulation of this model are presented, including the
 radiative flux and the total flux (the sum of radiative and convective fluxes).
 For comparison, the 1D (static) radiative flux is presented, too. 

\begin{figure}
\plotone{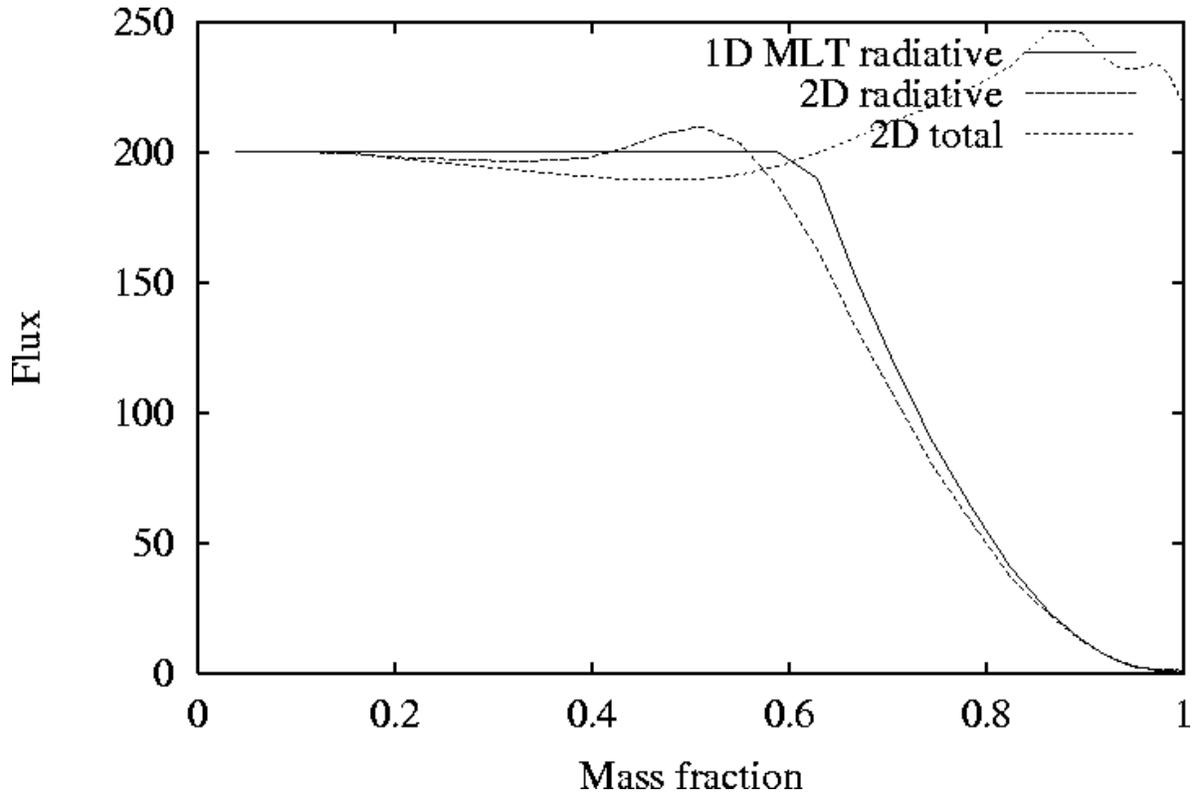}
\caption{Energy fluxes in the $\lambda = 1$ standard simulation in $L_{\sun}$ as a 
function
of the mass fraction in the simulated envelope. The outer boundary corresponds to
mass of $1.2 M_{\sun}$, and the total mass of the simulated region is $.013
 M_{\sun}$. The different flux components are indicated. \label{fig3}}
\end{figure}

\placefigure{fig3}

The radiative flux in the 2D simulation resembled the static model
, though a modest widening of the convective zone is clear, and an increase in 
the radiative flux above the constant luminosity is apparent near the bottom of
 the convective zone (see \cite{zn91}). The value of the total flux is indicative 
of the thermal equilibrium of the model. The total flux was slightly less than $200 
L_{\sun}$ below the convective zone, and reached a maximum of about $245
 L_{\sun}$ near the upper boundary and then it declined to the surface 
luminosity of about $215 L_{\sun}$. While it is clear that this is a 
result of a yet non equilibrium configuration, one shouldn't forget that due to 
the lack of algebric energy conservation, this plot is influenced by the 
uncertainty in the value of the hydrodynamical flux (as was described in \S 2):
 The energy conservation in each time step was of the order of $10^{-9}$ of the 
energy. This seems to be small, but when one divides it by the time step (about 
80 seconds), one gets $10 L_{\sun}$. I think it is important to note that  this 
has little effect upon the simulation itself, which I confirmed by
adding artificial non-energy-conserving term to the {\it 1D} dynamic 
code and comparing the results of several simulations where I changed the value of 
this term.

The different values of the luminosity in the simulations (figure 1) were 
dominated by the values of of the outer radii (compare this figure with figure 3
in AT97). In the initial configurations there was a relatively large difference 
in the effective temperatures: 3675K for $\lambda = 0$, 4600K for $\lambda = 
1$ and 5110K for $\lambda = 2$. However, in the dynamical 
simulations, very quickly (less than a month) the difference shrank 
dramatically, and the effective temperature became: 4520K, 4600K, and 4650K 
(for the same models as above) with time fluctuations of about 100K. In other 
words, using the dynamical simulations, one gets almost immediately, an 
effective temperature which is very close to the effective temperature of the 
$\lambda = 1$ model. This gives more confidence in the above conclusion that
 the $\lambda = 1$ is the preferred model. The 1D consistency 
simulations featured the same phenomenon.

\section{results of numerical parameter survey}

The five (2D) simulations used similar numerical 
parameters. The grid consisted of 36 angular cells in each row, with 53 to 72 
rows (for the simulations of models with $\lambda > 1$, I included  
deeper regions of the envelope - down to eight $R_{\sun}$ - because of the 
deeper convection zone). The angular sector was $0.2 \pi$ near the equatorial 
plane. The fluxes in the relaxation stage were calculated using a second order
donor scheme, and the energy flux consisted of both internal and kinetic 
energies. I used reflective boundary conditions on the sides of the 
computational domain. LES viscosity was used, with Smagorinsky coefficient of 
0.5 and without entropy diffusion.

I made many tests of the dynamical simulation of the model that started with the 
$\lambda = 1$ profile. I changed the numerical scheme, the number of cells 
and the angular width of the computational domain and the parameters of the LES 
terms. 

The changes in numerical scheme included: calculating the remapping fluxes using
 full donor (i.e. without taking the gradients of the hydrodynamical variables 
into account), using internal energy fluxes in the grid relaxation stage 
(without adding the kinetic energy) and the use of periodic
 boundary conditions at the sides. I found that the full donor simulation 
yielded a rapid expansion of the envelope - implying it is necessary to include 
gradients in the calculation of the fluxes. The use of the internal energy (only) 
for 
the remapping energy fluxes caused a small constant loss of kinetic energy and 
thus a modest shrinking of the envelope without changing the luminosity. The 
velocity field of the simulation with periodic boundary conditions is presented 
in figure 4. The non radial velocities at the sides are expected, but no other 
significant change can be seen, compared to the velocity field of reflecting 
boundary conditions (figure 2), i.e. the size of the eddies and the changes in 
the velocities with depth look similar. The luminosity and outer radius as a 
function of time look the same as well, as can be seen in figure 5.

\begin{figure}
\plotone{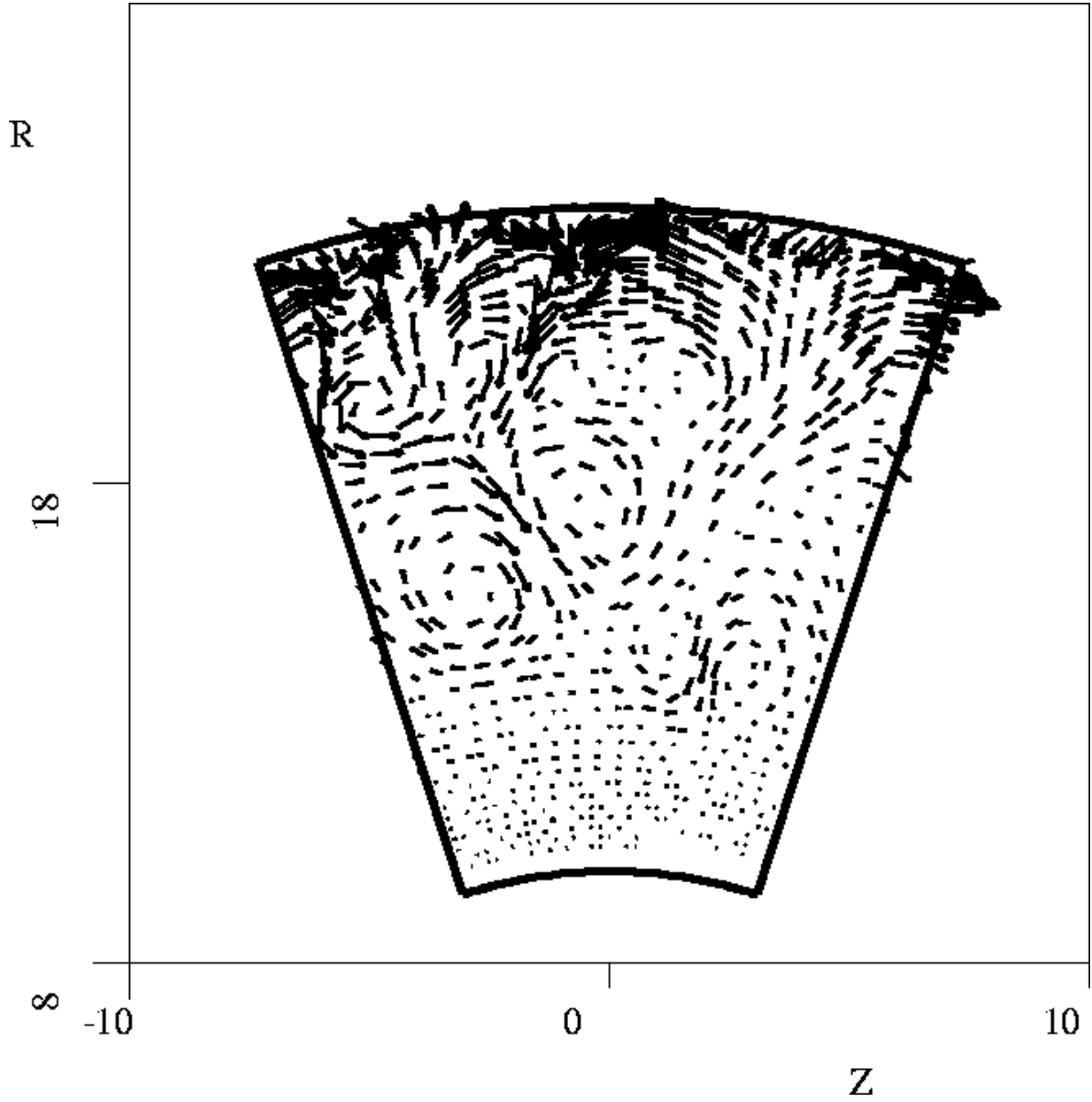}
\caption{The velocity field in the periodic boundary simulation. Z is the 
rotation axis and R is the distance from it in $R_{\sun}$,
 the largest arrows correspond to velocity of $10^{6} \case{cm}{s}$ \label{fig4}}
\end{figure}

\begin{figure}
\plottwo{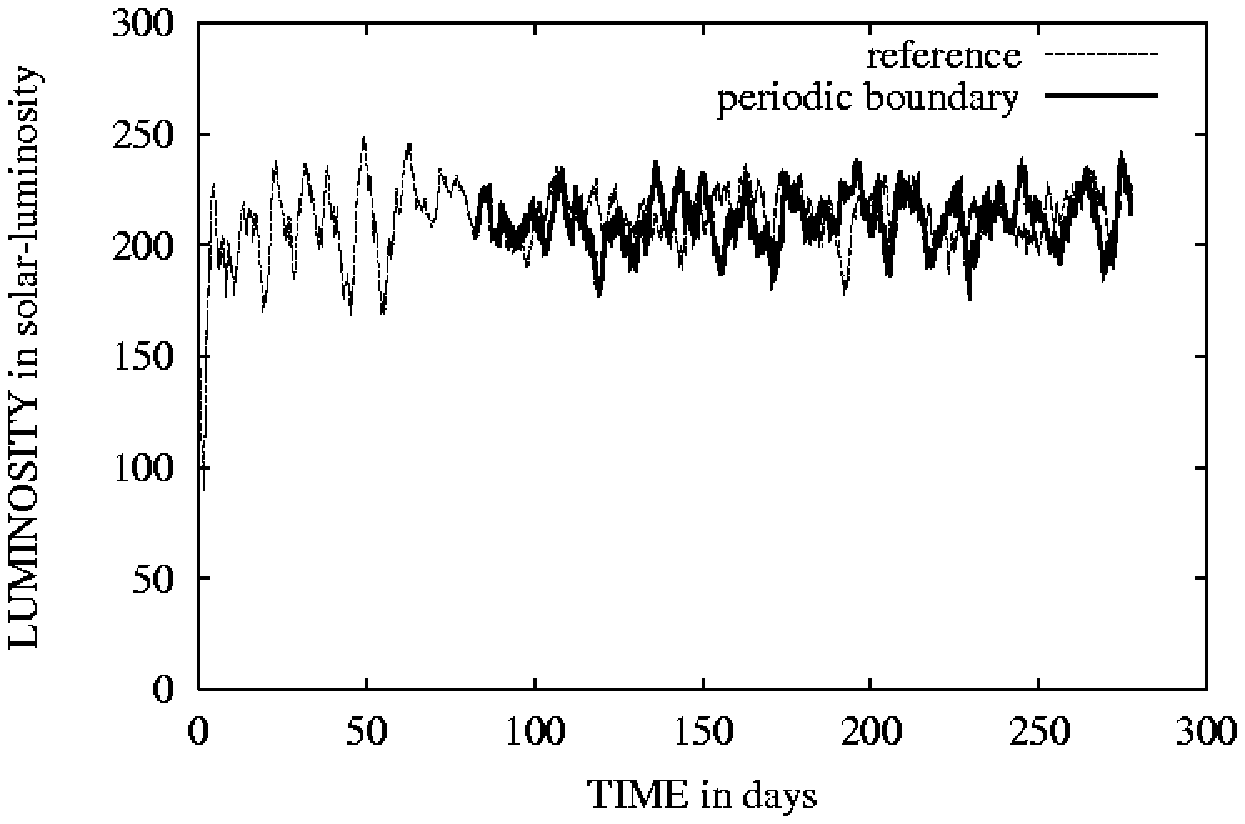}{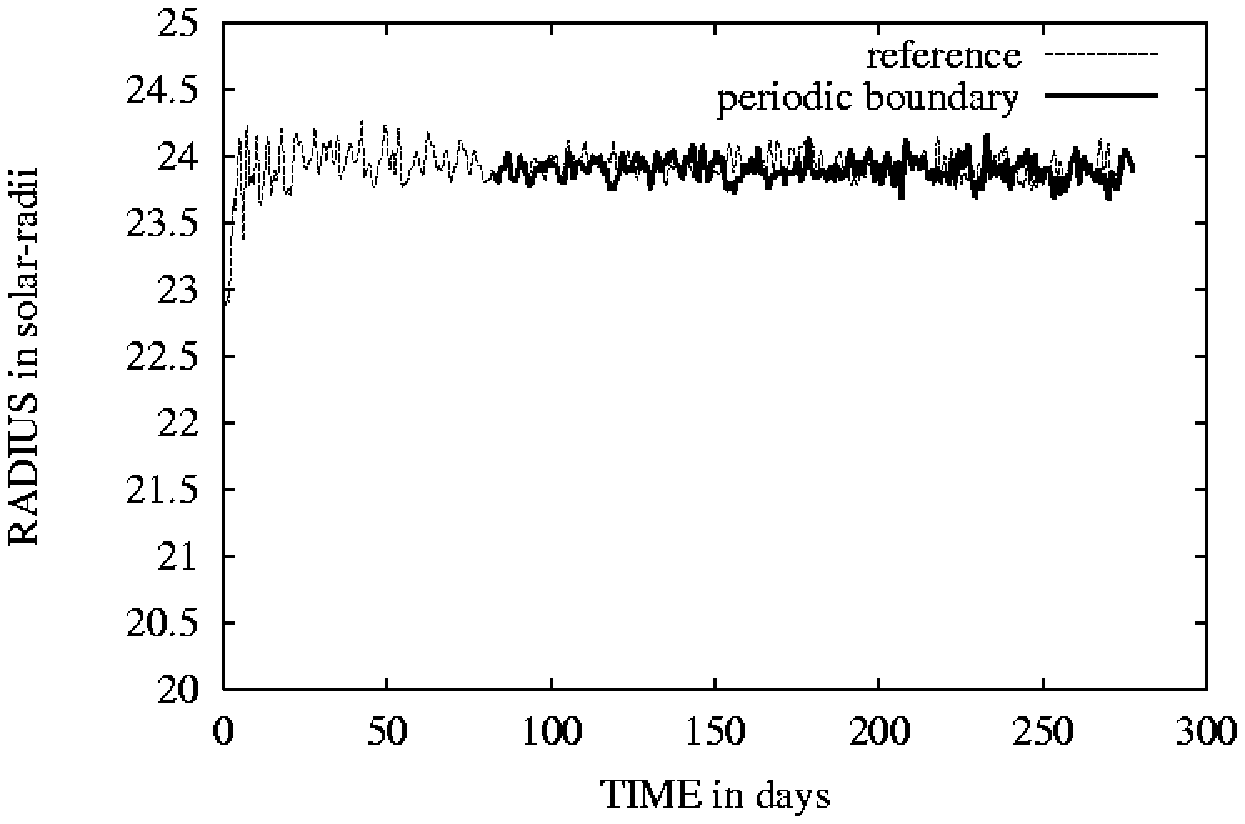}
\caption{(a) Luminosities ($L_{\sun}$) and (b) outer radii ($R_{\sun}$)  as a 
function of time (days) in the 2D simulation with periodic boundary (bold) and 
in the standard simulation (dash). \label{fig5}}
\end{figure}

\placefigure{fig4}
\placefigure{fig5}

Changes in the angular width or number of cells had small effects. Increasing 
both the angular width to $0.3 \pi$ and the number of cells in each row to 54 
yielded similar results as the nominal simulation, as presented in the velocity 
field - figure 6, and in the luminosity and outer radius - figure 7. When I 
decreased the angular width to $0.1 \pi$ I got a larger variability in the 
luminosity and outer radius (see figure 8) and a small increase in the average 
luminosity, as well. A small increase in the average luminosity was seen in a 
simulation with larger number of cells (=54) in each row (with the nominal 
angular width of $0.2 \pi$), as presented in figure 9. Increasing the number of 
rows to 81, had no significant effect.

\begin{figure}
\plotone{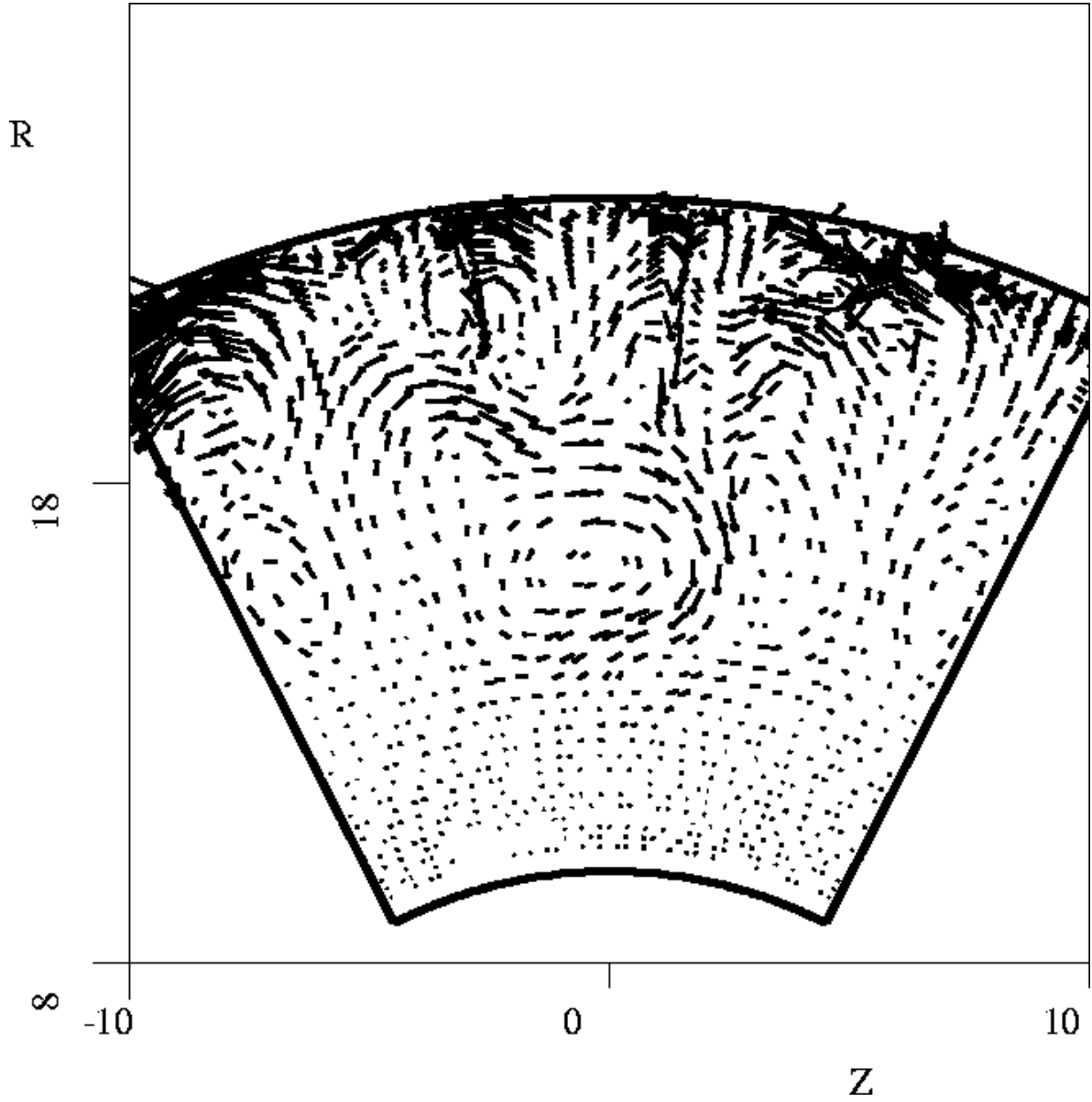}
\caption{The velocity field in the wider sector simulation. Z is the 
rotation axis and R is the distance from it in $R_{\sun}$,
 the largest arrows correspond to velocity of $10^{6} \case{cm}{s}$ \label{fig6}}
\end{figure}

\begin{figure}
\plottwo{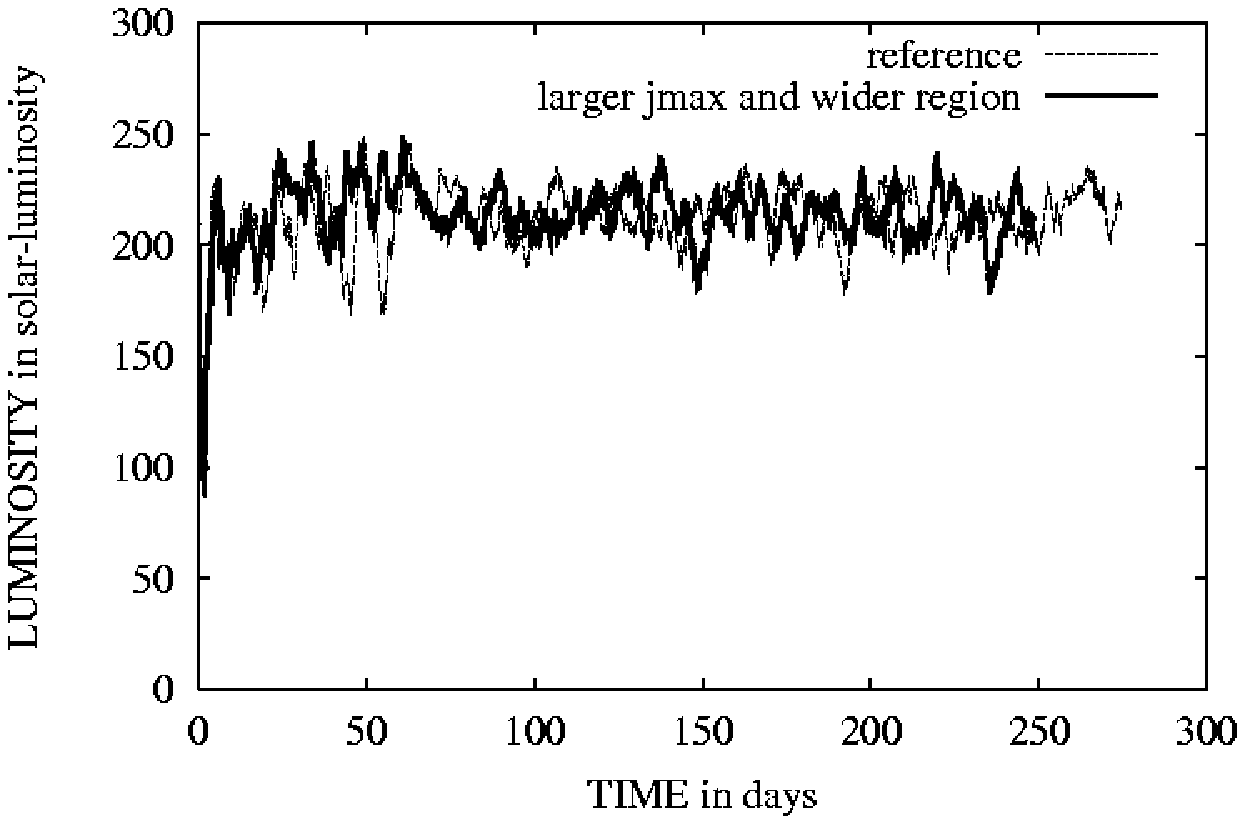}{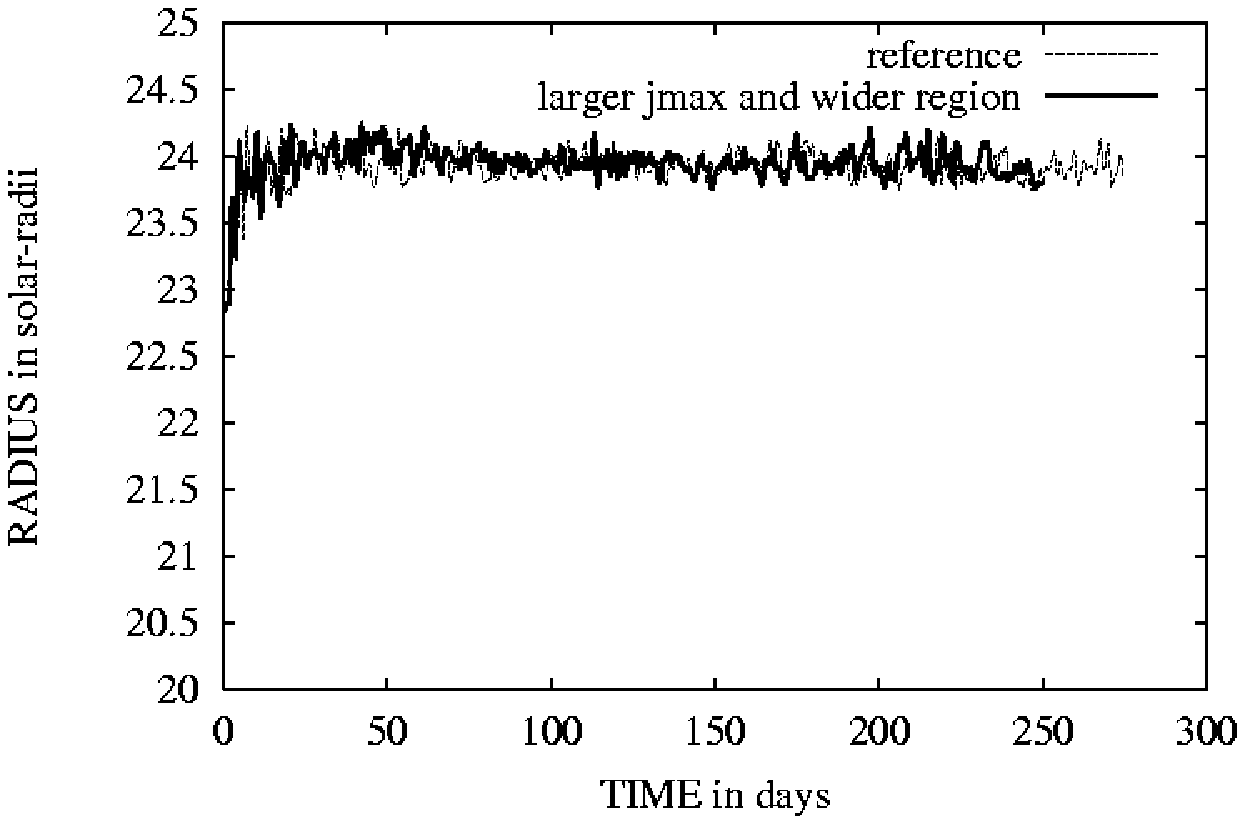}
\caption{(a) Luminosities ($L_{\sun}$) and (b) outer radii ($R_{\sun}$)  as a 
function of time (days) in the 2D simulation with a wider sector (bold) and 
in the standard simulation (dash). \label{fig7}}
\end{figure}

\begin{figure}
\plottwo{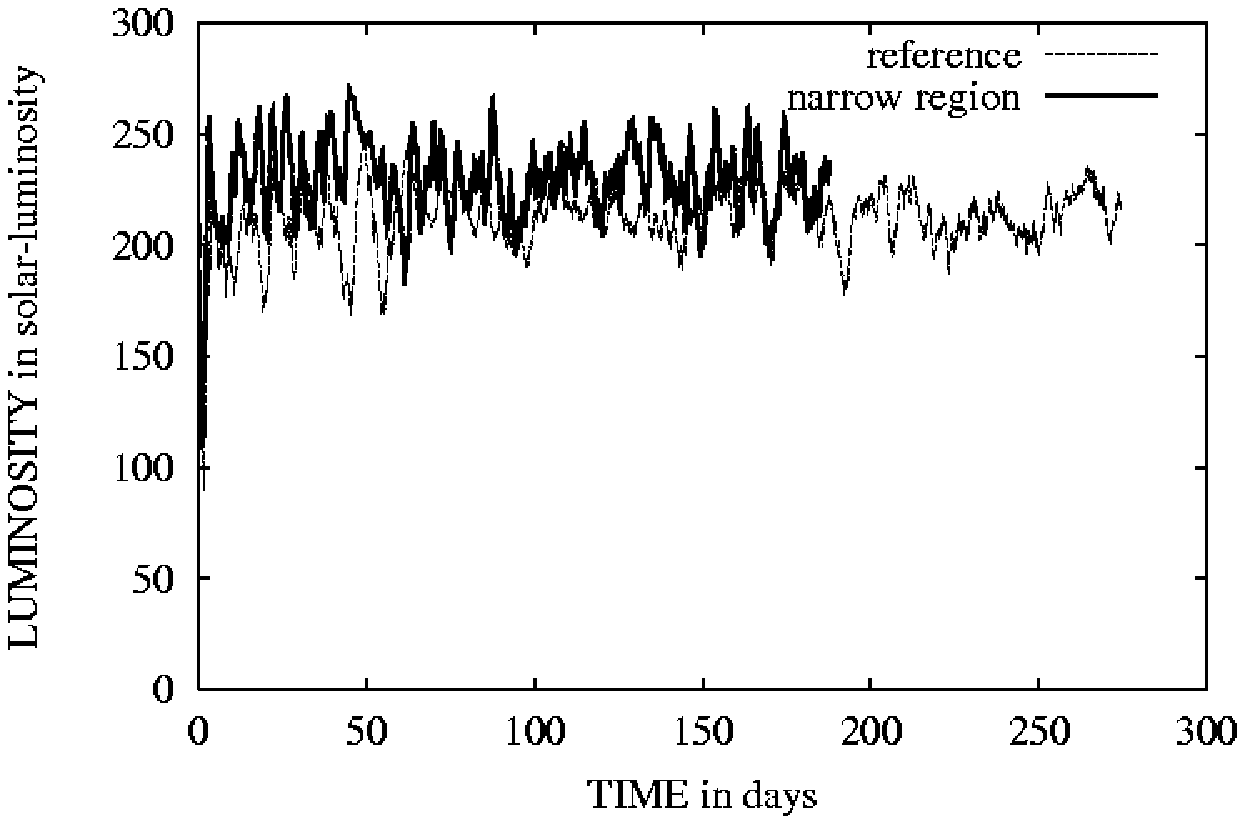}{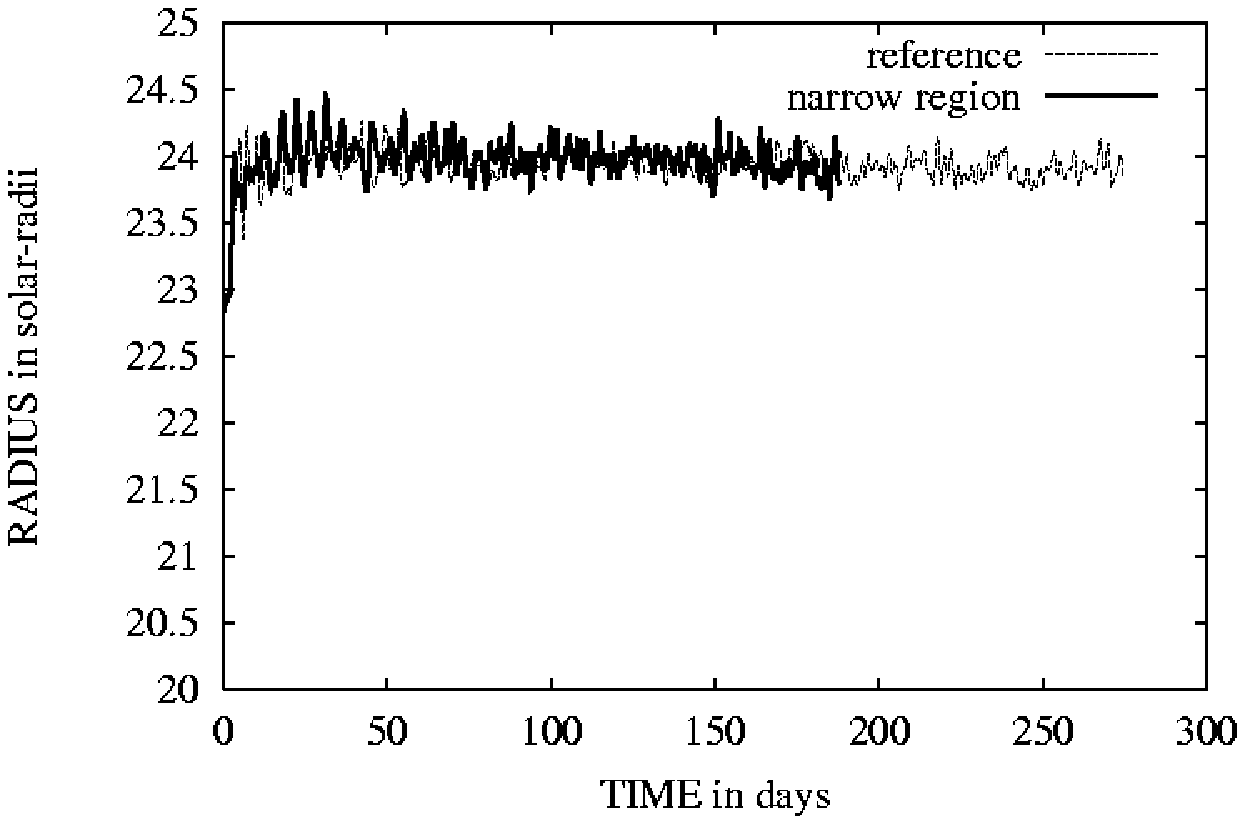}
\caption{(a) Luminosities ($L_{\sun}$) and (b) outer radii ($R_{\sun}$)  as a 
function of time (days) in the 2D simulation with a narrow sector (bold) and 
in the standard simulation (dash). \label{fig8}}
\end{figure}

\begin{figure}
\plottwo{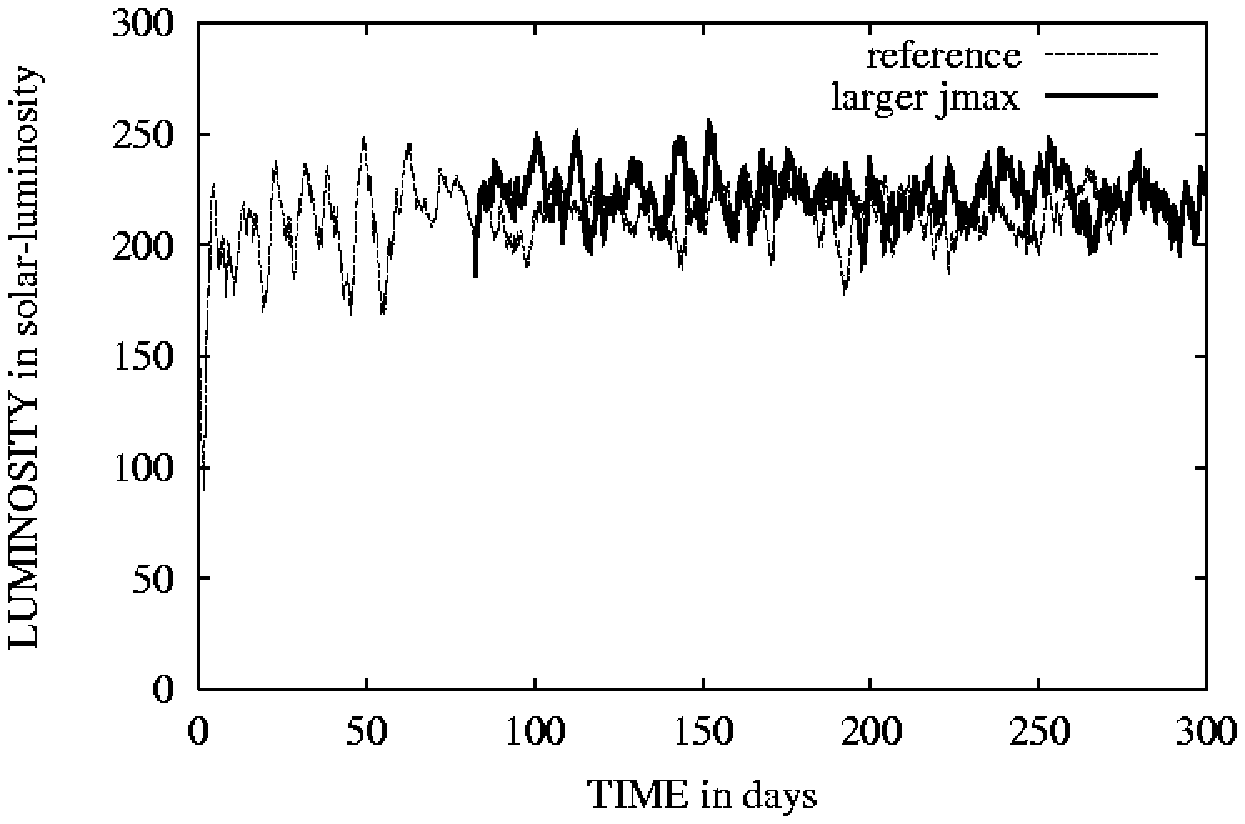}{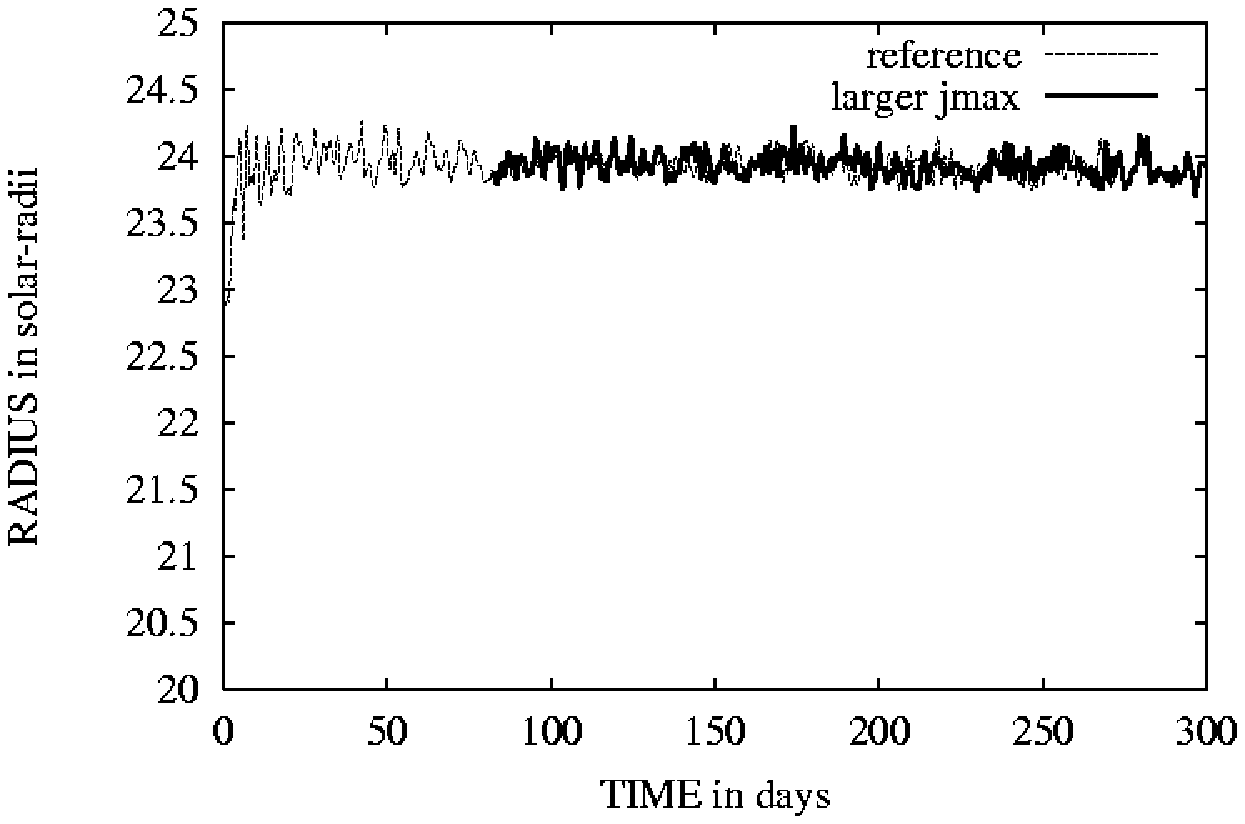}
\caption{(a) Luminosities ($L_{\sun}$) and (b) outer radii ($R_{\sun}$)  as a 
function of time (days) in the 2D simulation with more cells in each row (bold) 
and in the standard simulation (dash). \label{fig9}}
\end{figure}

\placefigure{fig6}
\placefigure{fig7}
\placefigure{fig8}
\placefigure{fig9}

The variations in the LES terms included changing the numerical coefficient and 
adding an entropy diffusion term. When I had zero or small (Smagorinsky 
coefficient of 0.01) LES viscosity I got somewhat larger variability in the 
luminosity and outer radius, but without changes of the average values. However,
including an entropy diffusion term (with Prandtl number of one) increased the 
average luminosity by a small amount (see figure 10). When I had this term, a 
further increase in the luminosity was evident when the Smagorinsky coefficient
 was increased to one.

\begin{figure}
\plottwo{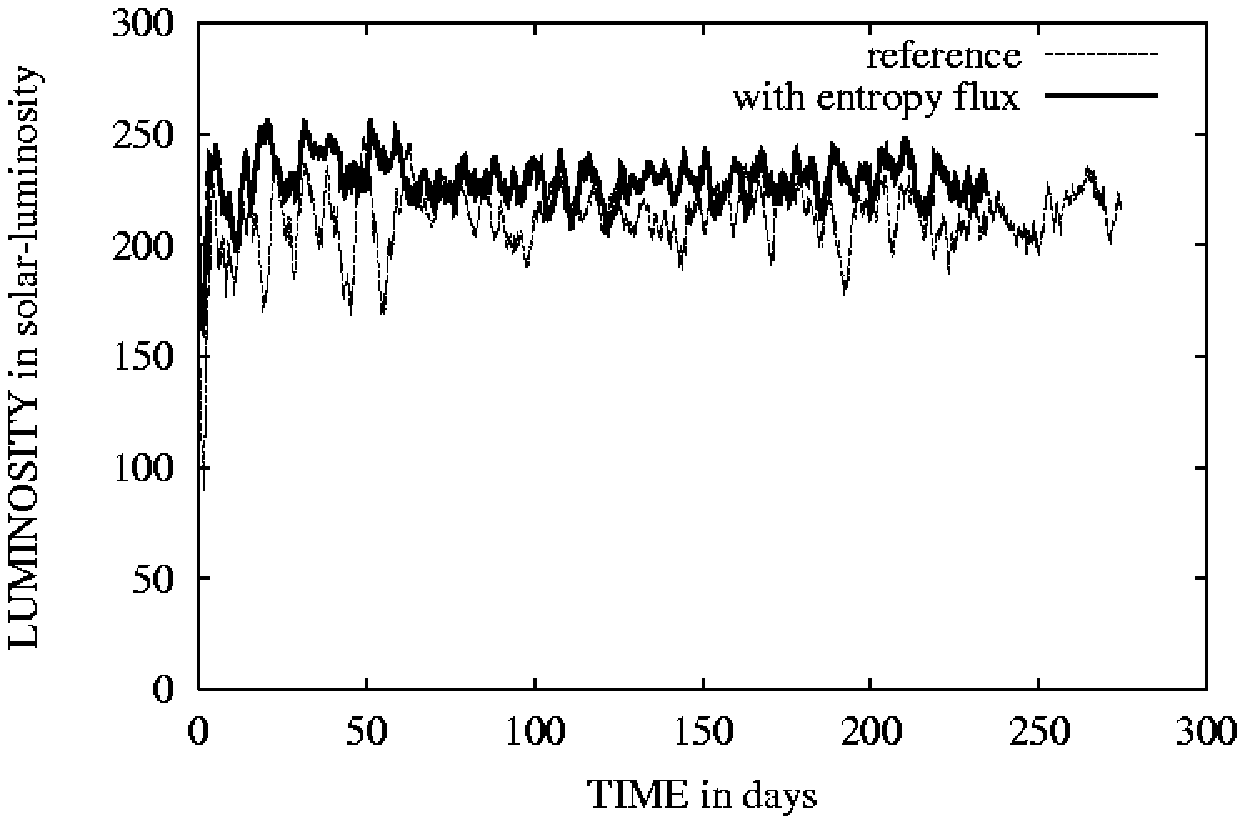}{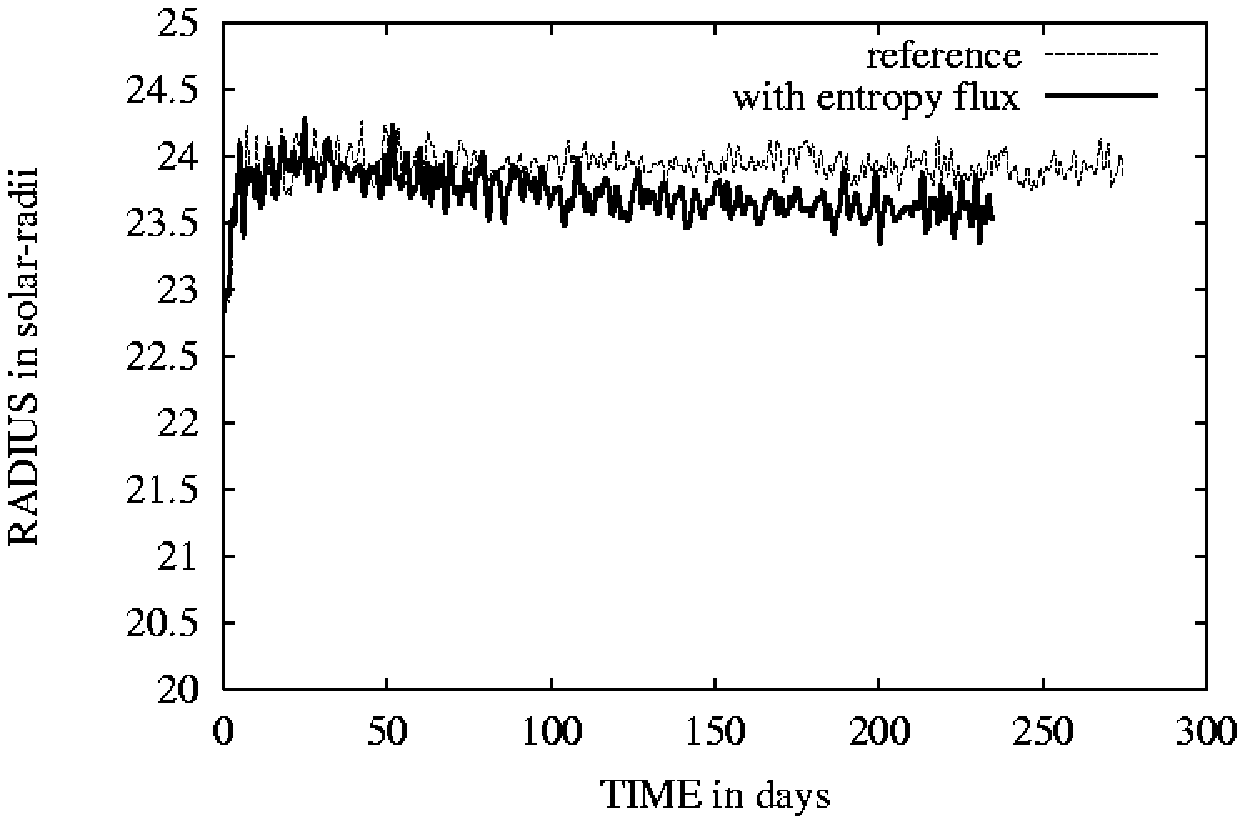}
\caption{(a) Luminosities ($L_{\sun}$) and (b) outer radii ($R_{\sun}$)  as a 
function of time (days) in the 2D simulation with LES entropy diffusion term (bold)
 and in the standard simulation (dash). \label{fig10}}
\end{figure}

\placefigure{fig10}

A summary of these tests reveals that there was a tendency to increase the 
luminosity in two cases: when I increased the angular resolution, and when I  
added an entropy diffusion term in the LES scheme. These two cases have one 
physical meaning: an increase in the efficiency of convection by adding 
smaller scale turbulent fluxes. Since there was a very small luminosity increase
in the above simulations, I performed yet another simulation where I increased the 
resolution by a (relatively) large factor. I started with a profile of the nominal 
simulation (of $36 \times 53$ cells) and increased the number of cells almost 
by an order of magnitude ($108 \times 149$ cells), without the entropy diffusion 
term. This simulation, 
being much more CPU demanding, was followed for a small physical time but 
nevertheless a clear conclusion can be made, as can be seen in figure 11. The 
luminosity increased quite significantly from about $215 L_{\sun}$ in 
the nominal simulation to about $250 L_{\sun}$.

\begin{figure}
\plotone{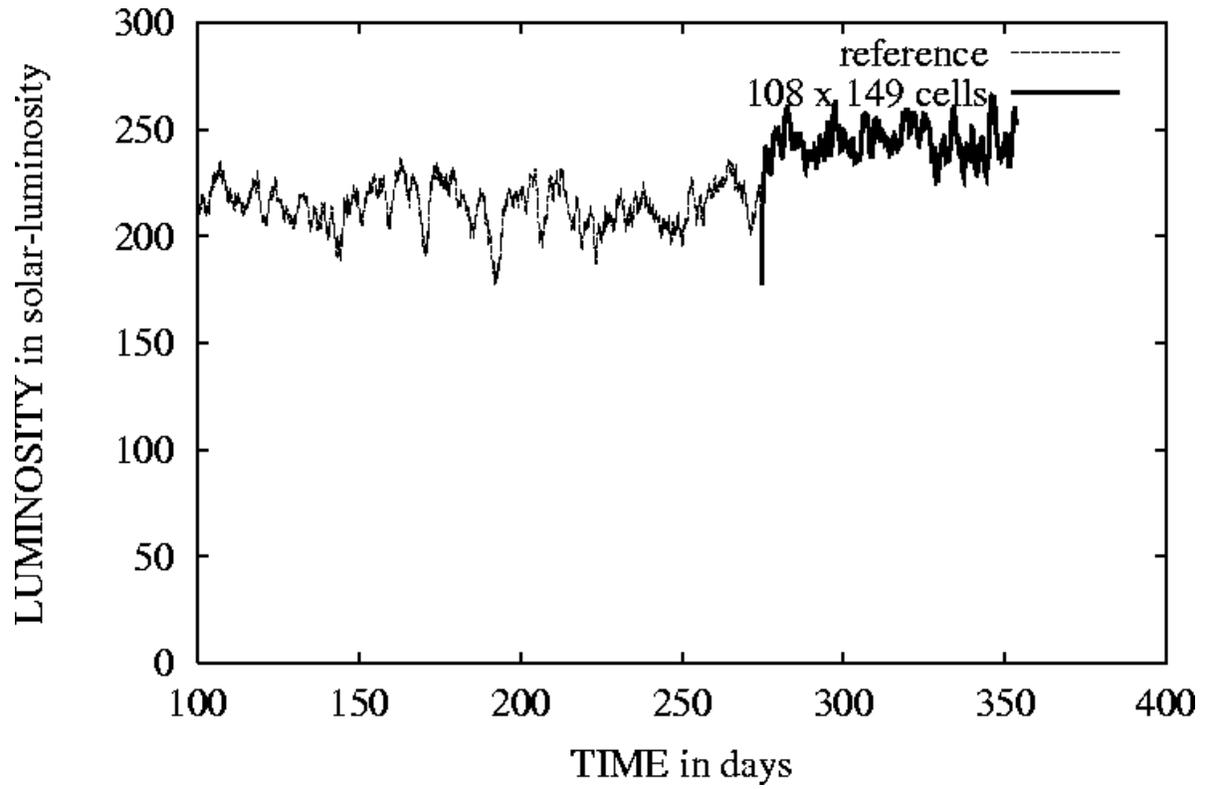}
\caption{Luminosities ($L_{\sun}$) as a 
function of time (days) in the 2D simulation with $108 \times 149$ cells (bold) and
in the standard simulation (dash). \label{fig11}}
\end{figure}

\placefigure{fig11}

An increase in luminosity means that the envelope loses more energy, and from the
five simulation with different initial models, this means that the 1D 
model which is consistent with the 2D simulations 
should have a larger value of $\lambda$, i.e. $\lambda > 1$. To test this
 conclusion I performed high resolution simulation starting with the initial model 
based upon $\lambda = 2$. Again, I started with a profile of the nominal 
simulation (with $36 \times 72$ cells) and increased the resolution to 
$108 \times 199$ cells. This yielded an increase in the luminosity from about 
$140 L_{\sun}$ to about $160 L_{\sun}$, as expected.

\section{Systematic tests for the luminosity}

In order to check systematically the dependence of the luminosity in the 
spatial resolution and in the LES terms, I performed the following simulations,
 which started with the same initial profile. The initial profile was a 2D
profile with an already developed convective flow based on the 
initial 1D model with $\lambda = 1$. From this profile I 
interpolated four equivalent profiles with different spatial resolution: 
$18 \times 26$ cells, $36\times 53$ cells, $72\times 106$ cells and 
$144\times 212$ cells, thus having factor of about 64 between the number
 of cells in the finer grid and in the coarser grid. From 
the above discussion it is clear that the significant term in the LES scheme is 
the entropy diffusion term, so I used the nominal 0.5 value of the Smagorinsky 
coefficient, and changed the importance of the entropy diffusion term by 
changing the Prandtl number, using the values: $1, \onehalf, 
\onethird,\ {\rm and}\ \slantfrac{1}{5}$ (i.e. the last value corresponds to 
increasing the diffusion term by a factor of five compared to the first value).
 I used these values of the Prandtl number for each of the four resolutions, 
and thus having sixteen simulations.

The flow in four of these sixteen simulations is presented in figure 12, for 
different resolutions. The flow is presented by contours of the vorticity and 
indeed increasing the resolution yielded more small scale eddies (and higher 
values of vorticity).

\begin{figure}
\plottwo{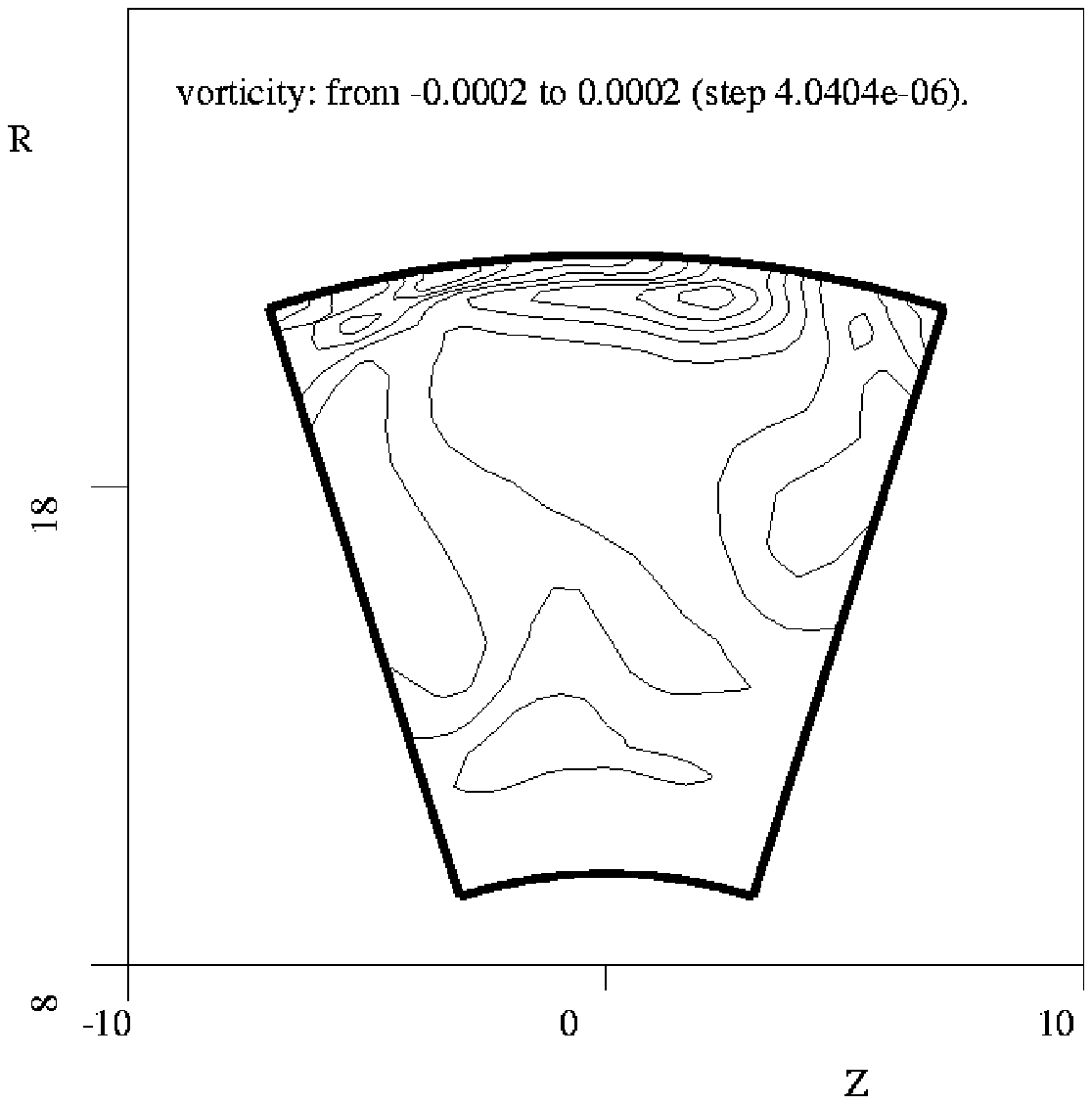}{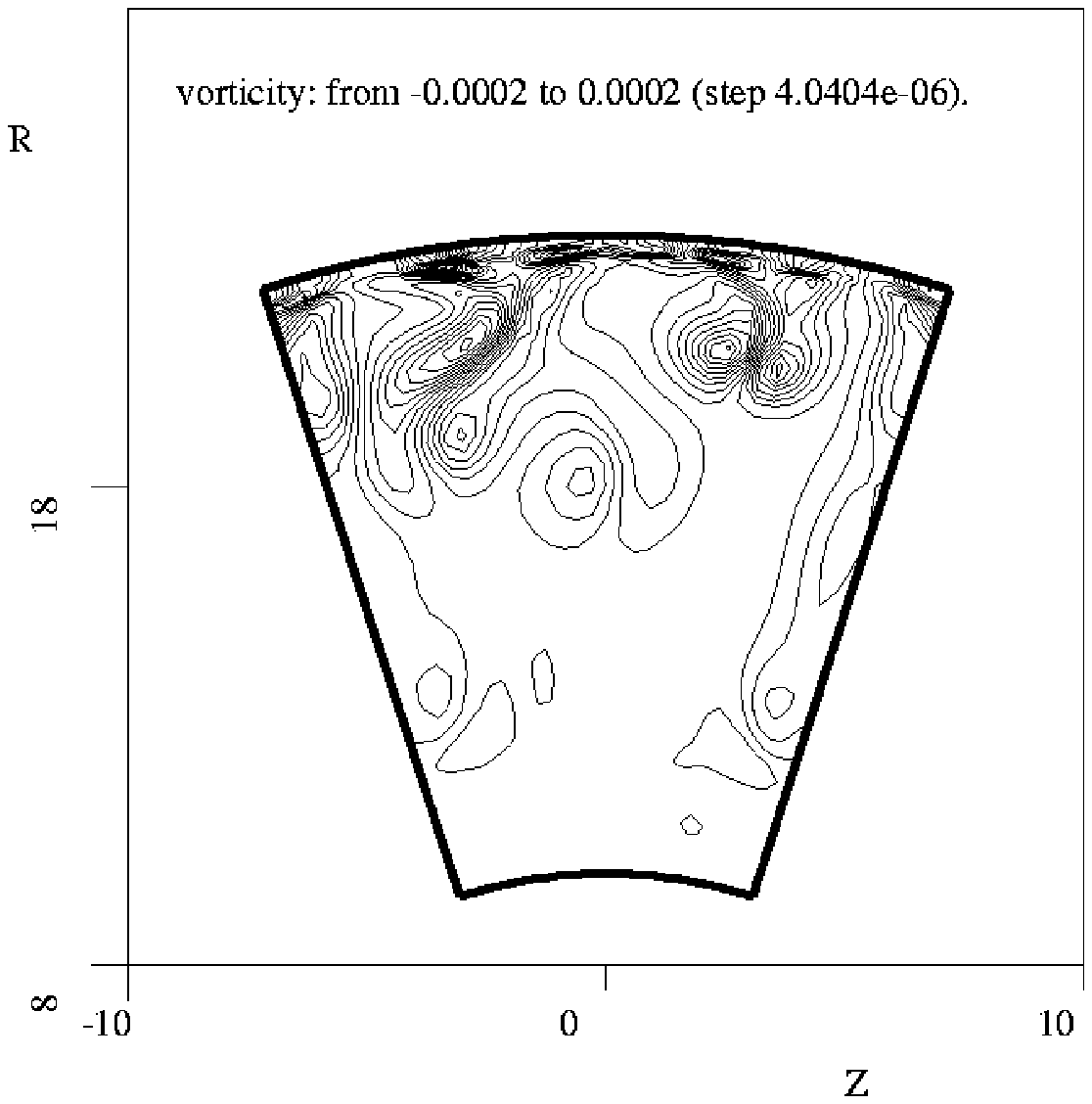}
\caption{Vorticity contours in the
(a)  $18 \times 26$ (b) $36 \times 53$ 
simulations. Z is the rotation axis and R is the distance from it in $R_{\sun}$.
\label{fig12}}
\end{figure}

\placefigure{fig12}

The luminosities of these simulations are summarized in figure 13, and in table 
1 I present the average luminosities. From these results one can conclude that 
the best estimate of the luminosity is $248\pm 2 L_{\sun}$, and that the
 better the resolution is, the less important the value of Prandtl number is. In
a perfect LES scheme one should get the converged result for each resolution for 
given set of numerical parameters (i.e. Prandtl number in our case). However, 
the value of about $248 L_{\sun}$ was obtained in each resolution with a
different value of Prandtl number, this value seems to increase when decreasing 
the resolution.

\placefigure{fig13}
\placetable{tbl-1}

For two of these 16 simulations I checked the importance of the LES entropy 
diffusion term comparing to the LES viscosity terms. I performed simulations where I
changed both Smagorinsky coefficient and Prandtl number so that the amplitude of
the diffusion term would be the same, while changing the amplitude of the 
viscosity terms. I performed a $72\times 106$ cell simulation with Smagorinsky 
coefficient of 1.5 and Prandtl number of 1 and got an average luminosity of 
$242 L_{\sun}$ (equivalent to a Prandtl number $\onethird$ simulation), and a 
$19\times 26$ cell simulation with Smagorinsky coefficient of 1. and Prandtl 
number of 1 where I got an average luminosity of $248 L_{\sun}$ (equivalent to 
a Prandtl number $\onehalf$ simulation).

The summary of these results is that the 2D simulations of an initial 
model calculated with $\lambda = 1$ MLT, yields an energy-losing envelope
with an average luminosity of $248 L_{\sun}$. To find what value of $\lambda$ 
can reproduce this I performed several 1D dynamic simulations. In
each of these simulations, I started with the same $\lambda = 1$ profile, 
and assuming different values for the $\lambda$ in the dynamical model I got 
different luminosities. For the following values of $\lambda$: 1.2, 1.3, 1.4 
and 1.5 I got average luminosities of about: 230, 240, 250 and 270 
($L_{\sun}$). From this I conclude that the consistent value of $\lambda$ 
is 1.4 .

\section{Summary}

As was presented in AT97 the convective envelope of a red giant  
(of mass $1.2 M_{\sun}$, core mass of $0.96 M_{\sun}$, luminosity of
$200 L_{\sun}$, and metallicity $Z=0.001$) was studied by 
comparing 2D simulations of five different initial models. While a 
full convergence of these simulations to a balanced final configuration was not 
possible, due to the large thermal time scale of this envelope, it was found that 
an initial model calculated with MLT using $\lambda = 1$, was almost 
energetically balanced, and thus was the best estimate of the final configuration. 

The fast change in the effective temperature obtained in these dynamic 
simulations, towards the effective temperature of the $\lambda = 1$  model is
 a further indication that this conclusion is consistent.

After a comprehensive survey of the importance of the various parameters used in 
the simulation technique, a sensitivity to the spatial resolution and to 
the LES entropy diffusion term was found. This sensitivity was tested by using four 
resolutions and four values of the LES term, and it was concluded that the 
preferred value of $\lambda$ is 1.4, for the envelope tested in this study. It was 
found that ,at least for this code and for these kind of simulations, one needs to 
calibrate the LES terms.

The conclusion for the value of $\lambda$, to be used in modeling the ged giant 
envelope studied in this research, being larger than one is consistent with the 
results of \cite{sc97} 
where they calculated evolutionary sequences and compared them to 
observation. This value of 1.4 is also consistent with {\it extrapolation} of the 
results of \cite{ldw99} and \cite{fls99} where numerical simulations were applied 
to the outer convective zone of solar like stars. 

This conclusion is important for studies in which an accurate modeling of Red 
Giants is necessary, like obtaining pulsation times of Miras (\cite{yt98}) 
or calculating isochrones of globular clusters(\cite{fs99}).

An obvious limitation of this study is the use of 2D simulations 
(and not 3D simulations). The spectrum of eddy sizes is probably different in
3D, but the influence of this upon the calibrated mixing length parameter is not
known. However, \cite{ldw99} have first estimation of the change in the calibrated 
mixing length parameter for their study of solar type convection, due to the 
difference between 2D and 3D simulations. They conclude that the calibrated mixing 
length parameter for the sun from 3D simulation is higher by 0.07 than from 2D 
simulation.

\acknowledgments
I am grateful to Yitzchak Tuchman for his encouraging cooperation and to Eli Livne 
for his code. I thank Ami Glasner, Yossi Stein and Zalman Barkat for many fruitful
discussions, and Dave Arnett and the referee for useful suggestions to the 
manuscript.

\clearpage

\begin{deluxetable}{ccccc}
\tablewidth{27pc}
\tablecaption{Average Luminosities ($L_{\sun}$) \label{tbl-1}}
\tablehead{
\colhead{$\frac{no.\ of\ cells}{Prandtl\ no.}$} &
\colhead{$18 \times 26$} &
\colhead{$36 \times 53$} &
\colhead{$72 \times 106$} &
\colhead{$144 \times 212$} 
}

\startdata
$1$                & 224 & 225 & 235 & 250 \nl
$\onehalf$         & 250 & 234 & 239 & 246 \nl
$\onethird$        & 272 & 247 & 242 & 248 \nl
$\slantfrac{1}{5}$ & 316 & 265 & 248 & 250 \nl
\enddata
\end{deluxetable}

\clearpage

\clearpage

\clearpage

\clearpage

\clearpage

\clearpage

\clearpage

\clearpage

\clearpage

\clearpage

\clearpage

\clearpage

\clearpage
\begin{figure}
\figurenum{12 -cont}
\plottwo{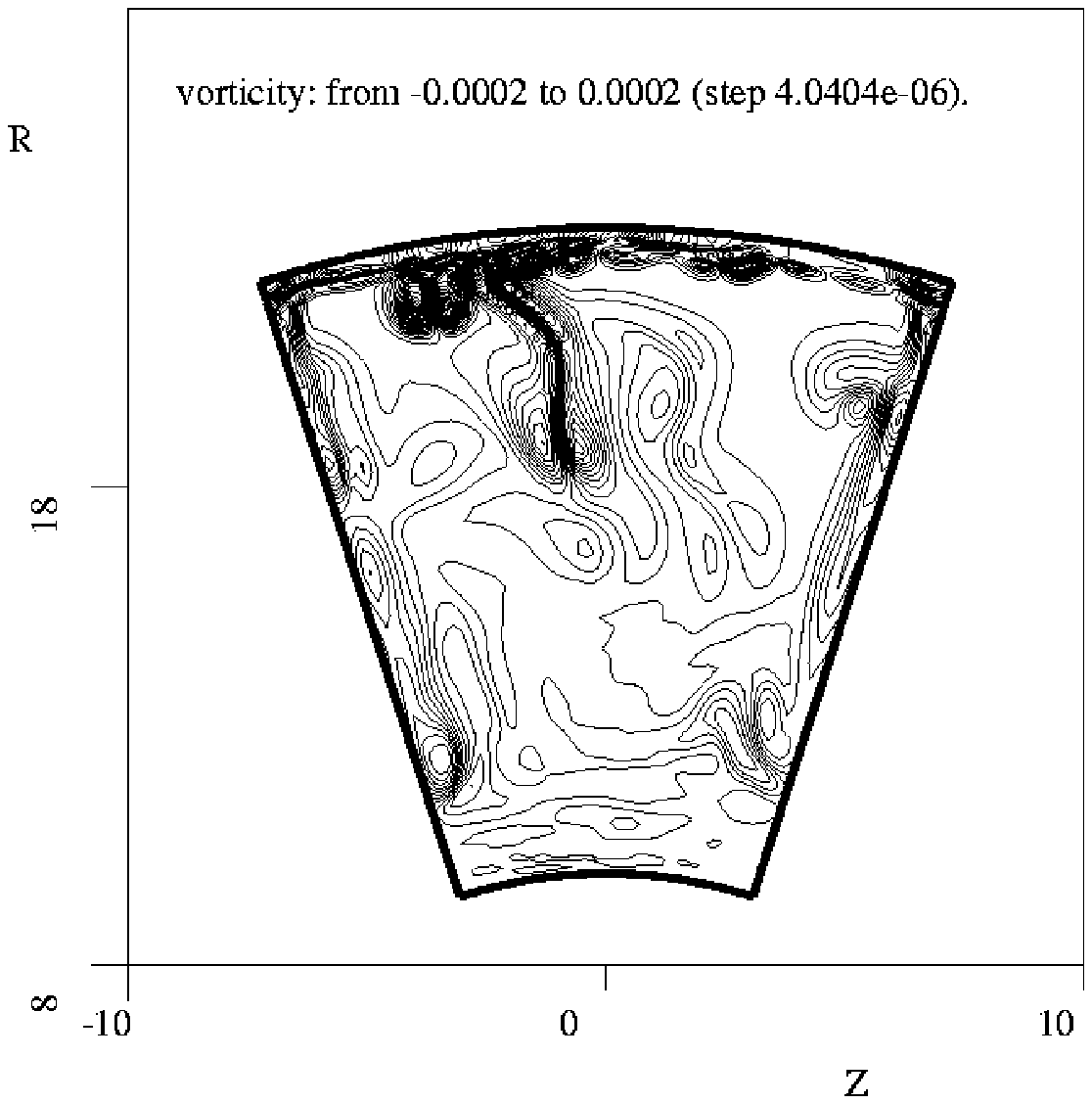}{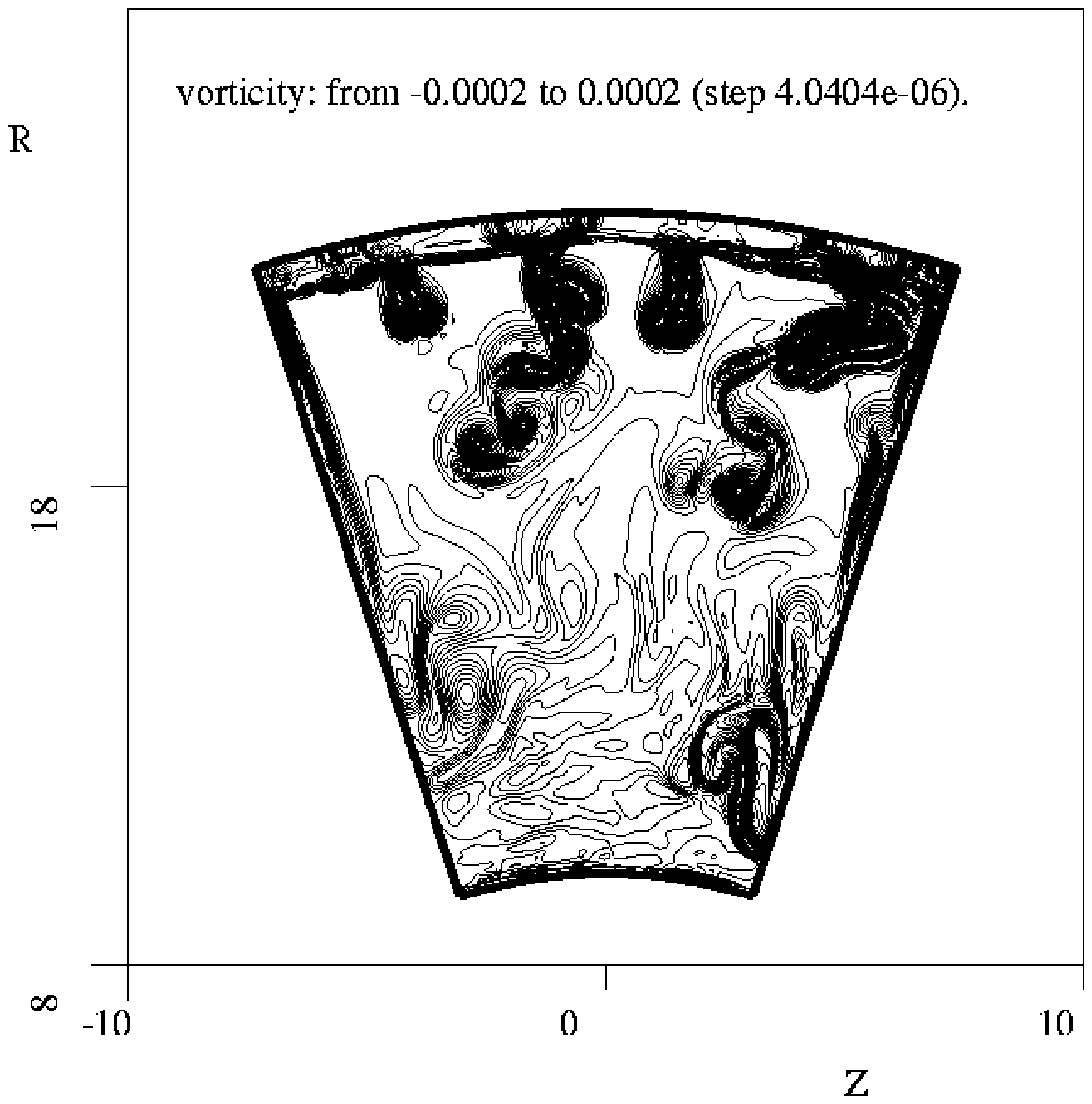}
\caption{Vorticity contours in the
(c) $72 \times 106$ and (d) $144 \times 212$
simulations. Z is the rotation axis and R is the distance from it in $R_{\sun}$.
}
\end{figure}

\clearpage
\begin{figure}
\plottwo{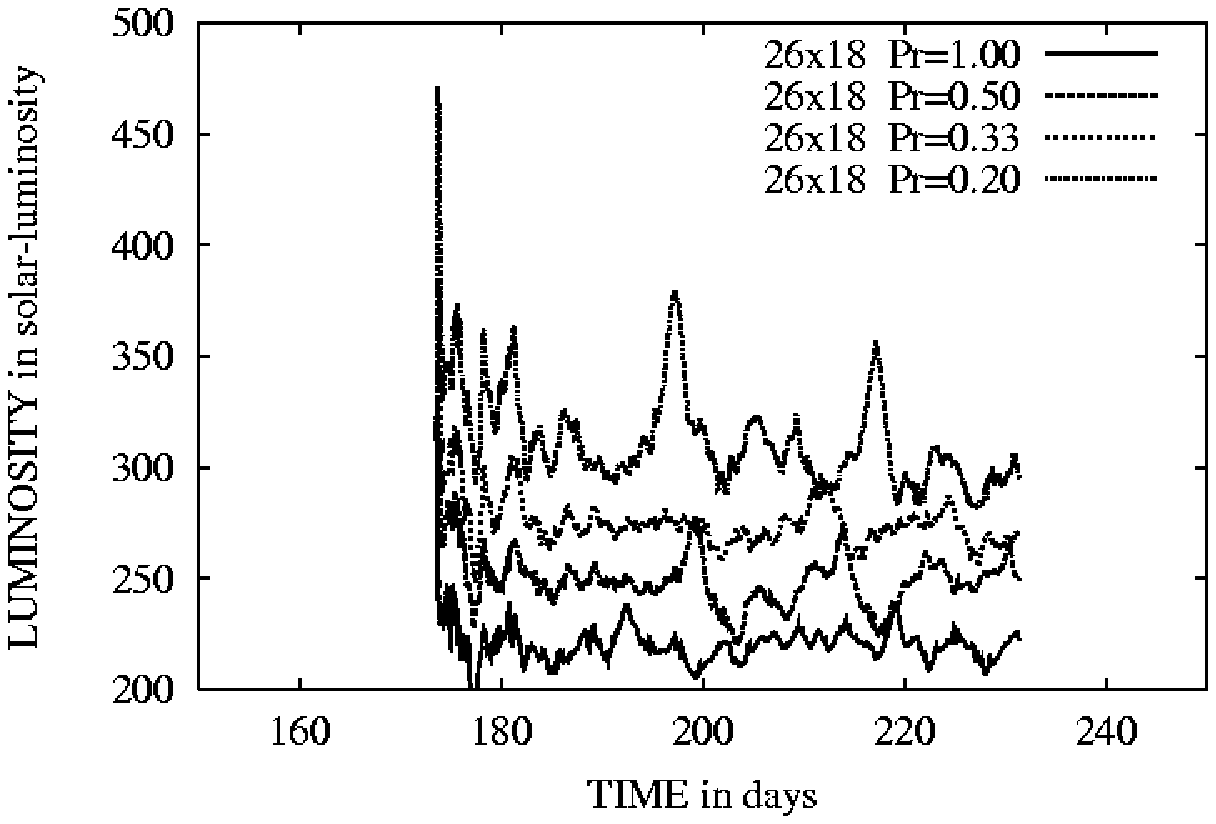}{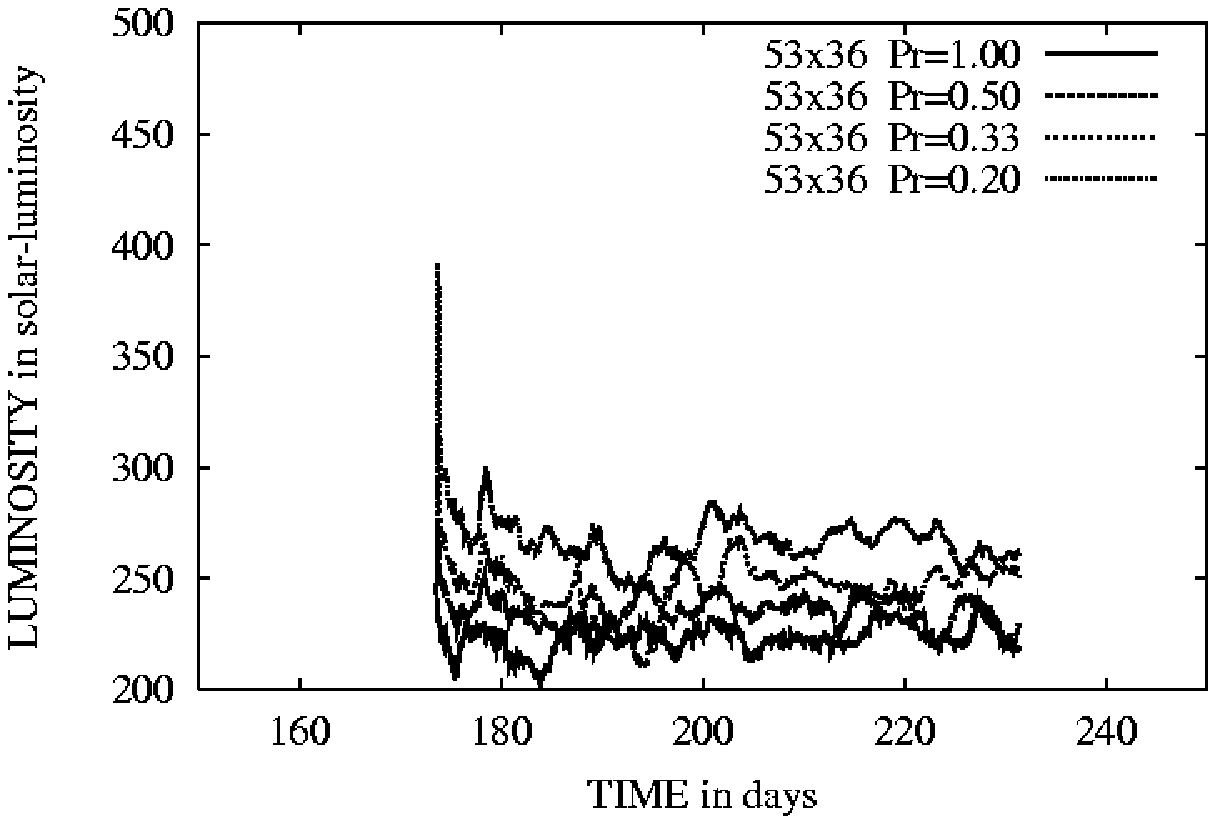}
\caption{Luminosities ($L_{\sun}$) 
as a function of time (days) in the 2D simulations with (a)  $18 \times 26$ 
(b) $36 \times 53$. In each
figure the four different Prandtl no. simulations are presented. \label{fig13}}
\end{figure}

\clearpage
\begin{figure}
\figurenum{13 -cont}
\plottwo{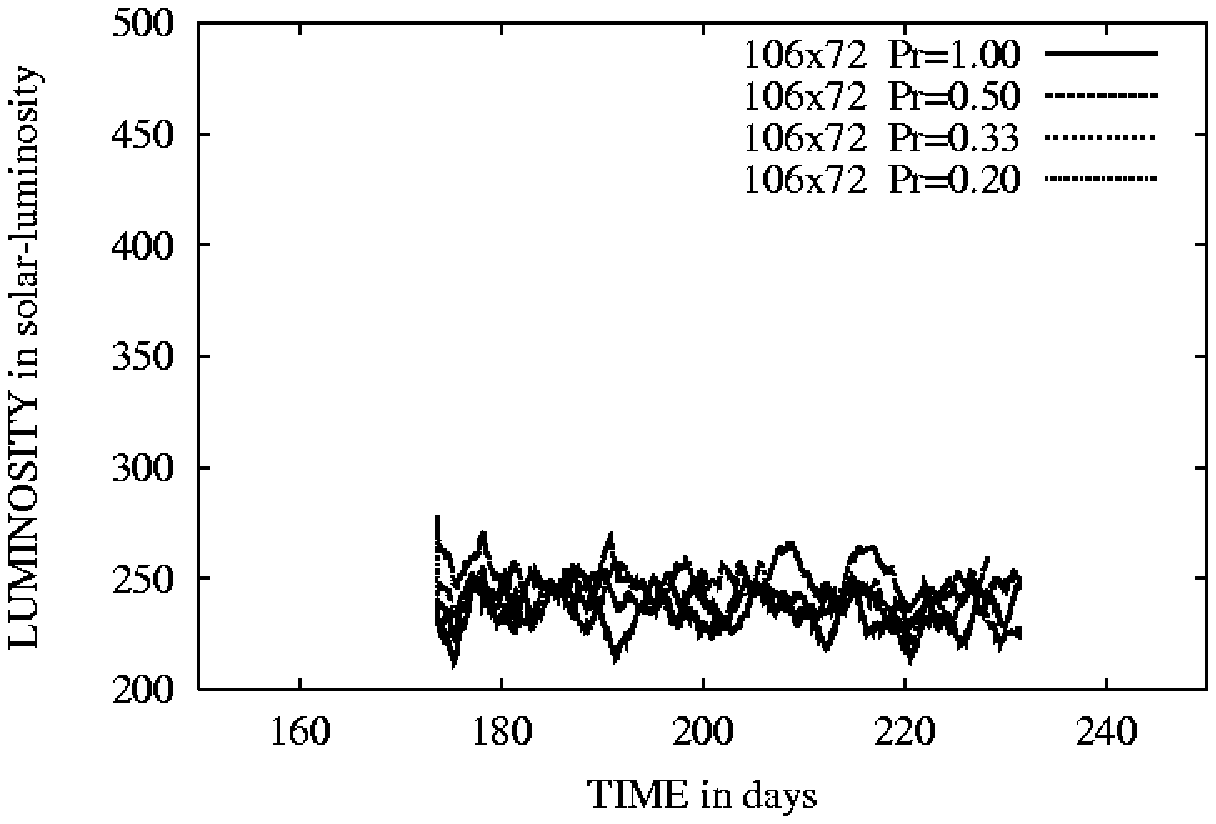}{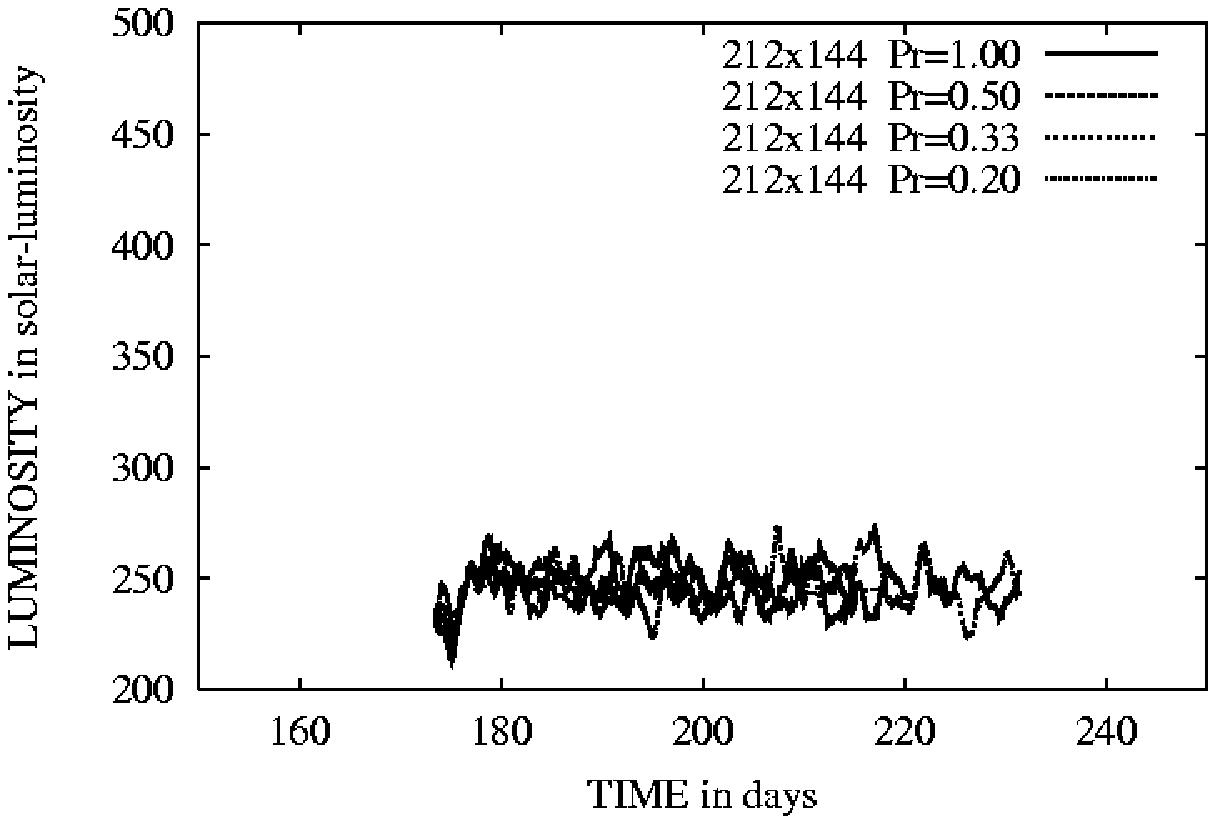}
\caption{Luminosities ($L_{\sun}$) 
as a function of time (days) in the 2D simulations with (c) $72 \times 106$ and 
(d) $144 \times 212$ cells. In each
figure the four different Prandtl no. simulations are presented. }
\end{figure}

\end{document}